\documentstyle[12pt,amsmath,amssymb,epic]{article}

\newcommand{\bbbone}{\mathchoice {\rm 1\mskip-4mu l} {\rm 1\mskip-4mu l}
{\rm 1\mskip-4.5mu l} {\rm 1\mskip-5mu l}}
\newcommand{\scalprod}[2]{\left\langle {#1}, {#2}\right\rangle}
\newcommand{\dom}{{\cal D}}
\newcommand{\RE}{{\rm Re}}
\newcommand{\IM}{{\rm Im}}
\newcommand{\fer}[1]{(\ref{#1})}
\newcommand{\ran}{{\rm Ran\,}}
\newcommand{\repsilonbar}{\,\overline{\!R}_\epsilon}
\newcommand{\repsilon}{R_\epsilon}

\newcommand{\h}{{\cal H}}
\newcommand{\cx}{{\mathbb C}}
\newcommand{\r}{{\mathbb R}}

\newcommand{\What}{{\widehat W}}

\renewcommand{\d}{{\rm d}}
\newcommand{\Pibar}{\overline{\Pi}}
\newcommand{\Pbar}{\overline{P}}

\newcommand{\w}{{\frak W}}
\newcommand{\alphaepsilon}{{\alpha_{t,\lambda}^{(\epsilon)}}}
\newcommand{\sigmaepsilon}{{\sigma_{t,\lambda}^{(\epsilon)}}}
\newcommand{\oref}{\omega^{\rm ref}}
\newcommand{\Oref}{\Omega^{\rm ref}}

\newcommand{\av}[1]{\left\langle{#1}\right\rangle}

\newcommand{\pomega}{P_{\Omega}}

\newcommand{\x}{\langle x \rangle}
\newcommand{\pbar}{\,\overline{p}\,}

\newcommand{\m}{{\frak M}}
\newcommand{\cc}{{\cal C}}

\newtheorem{theorem}{Theorem}[section]
\newtheorem{lemma}{Lemma}[section]
\newtheorem{proposition}{Proposition}[section]

\begin{document}

\title{Ionization of atoms in a thermal field}

\author{J. Fr\"ohlich\footnote{Institut f\"ur Theoretische Physik, 
ETH H\"onggerberg, CH-8093 Z\"urich, Switzerland (juerg@phys.ethz.ch)}
\and
M. Merkli\footnote{Institut f\"ur Theoretische Physik, 
ETH H\"onggerberg, CH-8093 Z\"urich, Switzerland; present address: Department of Mathematics and Statistics, McGill University, Montreal, PQ, H3A 2K6, Canada (merkli@math.mcgill.ca)}
\and
I.M. Sigal
\footnote{Department of Mathematics, University of Toronto, Toronto, 
ON, M5S 3G3 Canada (webpage: www.math.toronto.edu/sigal)}
\thanks{Supported by NSERC under grant NA7901.}}
\maketitle

\begin{center}
{\em This paper is dedicated to Elliott H. Lieb,\\
 in admiration and friendship.}
\end{center}

\begin{abstract}
We study the stationary states of a quantum mechanical system describing an
atom coupled to black-body radiation at positive temperature. The
stationary states of the non-interacting system are given by product states,
where the particle is in a  
bound state corresponding to an eigenvalue of the particle Hamiltonian, and
the field is in its equilibrium state. 
We show that if Fermi's Golden Rule predicts that a stationary state
disintegrates after coupling to the radiation field then it is unstable,
provided the coupling constant is sufficiently small (depending on the
temperature). \\
\indent
The result is proven by analyzing the spectrum of the thermal Hamiltonian
(Liouvillian) of the system within the framework of $W^*$-dynamical systems. A
key element of our spectral analysis is the positive commutator method.
\end{abstract}
{\bf Keywords\ } {\small Open quantum system, black-body radiation, CCR algebra,
positive temperature representation, Fermi Golden Rule, Liouville
operator, virial theorem, positive commutator method}

\thispagestyle{empty}
\newpage
\tableofcontents

\setcounter{section}{0}
\section{Introduction}

In this paper, we study a quantum mechanical model of an atom interacting
with black-body radiation. The atom is described by an electron moving in a
potential, e.g. the Coulomb potential of a static nucleus. \\
\indent
Our goal is to show that when the atom is coupled to the quantized radiation 
field in the state of black-body radiation at sufficiently high temperature,
it is ionized. We describe this ionization process by showing that 
stationary states of the system become unstable when the atom is coupled to
the radiation field. Each bound state of the atom leads 
to a stationary state of the uncoupled system, where the field is in an 
equilibrium state. We consider a class of small interactions localized in
space which couple {\it finitely} many bound states of the atom to the
field. We show that the stationary states for the coupled system correspond to those eigenvalues of the atomic Hamiltonian which
are not coupled to the field. In other words, all stationary states arising
from atomic bound states which are coupled to the field by the interaction
are unstable, provided the coupling constant is small enough. \\
\indent
This instability is explained by the following mechanism. If the electron is in a
bound state with energy $E<0$, it will be hit, after some time, by a quantum
(photon) of energy $\omega\geq -E$, and hence will make a transition to a scattering state of energy
$E+\omega$. In other words, the atom is ionized. \\
\indent
The average time, $t_E$, it takes for an atom in a bound state of energy $E$ to be
ionized is given, to second order in the perturbation, by the Fermi Golden
Rule. Heuristically, for an inverse temperature $\beta$ of the field, it
satisfies  
\begin{equation}
t_E\propto e^{\beta|E|} |p(E)|^{-2},
\label{te}
\end{equation}
where $|p(E)|^2$ is the probability for the electron to make a transition from
the bound state to a scattering state by absorbing a photon of energy $>E$. The factor
$e^{\beta|E|}$ in \fer{te} can be explained by Planck's law, which says that the
probability density for a photon to have energy $\omega$ is 
$\frac{1}{e^{\beta\omega}-1}$. At zero temperature, $\beta=\infty$, one finds
that $t_E=\infty$, and thermal ionization does not occur. For a given strength
of the interaction, as measured by the size of a coupling constant
$\lambda\in\r$, we are 
able to show that thermal ionization takes place,
provided $0<|\lambda|<ke^{-2\beta |E_0|}$, where $E_0<0$ is the minimal 
energy of the electron in a bound state coupled to the radiation field, and
$k$ is some constant. This restriction is of 
technical nature; physically, thermal ionization is expected to be observed for
arbitrarily small temperatures, provided the coupling constant is small enough
{\it independently} of $\beta$. \\

Next, we describe the system and our main results in some more detail. The
atomic Hamiltonian is a Schr\"odinger operator $H_p=-\Delta+v$ on the Hilbert
space 
$\h_p=L^2(\r^3,d^3x)$, where $v$ belongs to a certain class of potentials including
the Coulomb potential regularized at the origin. The operator $H_p$ generates the
Heisenberg dynamics
\begin{equation*}
\alpha_t^p(A) =e^{it H_p} Ae^{-itH_p}
\end{equation*}
on the von Neumann algebra ${\frak A}_p={\cal B}(\h_p)$ of bounded operators
on $\h_p$. \\
\indent
The field is conveniently described in terms of a $C^*$-algebra ${\frak A}_f$, which can
be viewed as a time-averaged Weyl algebra. The dynamics is given by a
$*$auto\-morphism group of ${\frak A}_f$ describing free massless bosons.\\
\indent
The combinded system is described in terms of the algebra 
\begin{equation*}
{\frak A}={\frak A}_p\otimes{\frak A}_f,
\end{equation*}
and the uncoupled dynamics is given by the automorphisms
\begin{equation*}
\alpha_{t,0}=\alpha_t^p\otimes \alpha_t^f.
\end{equation*}
To define the coupled dynamics, we specify a (regularized) interaction term
$\lambda V^{(\epsilon)}$, whose form is motivated by standard models of atoms
interacting with the radiation field. The regularization is introduced to
guarantee that $V^{(\epsilon)}\in\frak A$, for all $\epsilon\neq 0$. The
interacting dynamics, $\alpha_{t,\lambda}^{(\epsilon)}$, is then defined as the
$*$automorphism group of $\frak A$ obtained by the 
Schwinger-Dyson series. \\
\indent
At zero temperature, the dynamics of the model is generated by the formal
Hamiltonian
\begin{equation*}
H=H_p+H_f+\lambda V,
\end{equation*}
where $H_f=\d\Gamma(|k|)$ is the free-field Hamiltonian, i.e., the second
quantized multiplication operator $|k|$, acting on bosonic Fock space ${\cal
  F}(L^2(\r^3,d^3k))$, and the interaction term $V$ is given by
\begin{equation*}
V=\sum_\alpha G_\alpha\otimes \left( a(g_\alpha)+a^*(g_\alpha)\right).
\end{equation*}
The sum is over a finite set, $G_\alpha$ are bounded selfadjoint operators on
${\cal B}(\h_p)$, and the form factors $g_\alpha$ are in $L^2(\r^3,d^3k)$.

\indent
We introduce a reference state
\begin{equation*}
\omega^{\rm ref} =\omega^p\otimes\omega_\beta^f
\end{equation*}
on $\frak A$, where $\omega^p$ is given by a (strictly positive) density
matrix, and $\omega_\beta^f$ is the $(\beta,\alpha_t^f)$-KMS state of ${\frak
  A}_f$, i.e., the state of black-body radiation at inverse temperature
$\beta$. We are interested in the time evolution of states on $\frak A$ which
are close to (normal w.r.t.) $\omega^{\rm ref}$, i.e., which are represented
by a density matrix on the GNS Hilbert space $\h$ of $({\frak
  A},\omega^{\rm  ref})$. The GNS representation provides us with a
representation map $\pi_\beta:{\frak A}\rightarrow {\cal B}(\h)$ and a vector
$\Omega^{\rm ref}\in \h$ s.t. $\omega^{\rm ref}(A)=\scalprod{\Omega^{\rm
    ref}}{ \pi_\beta(A)\Omega^{\rm ref}}$. There is a selfadjoint operator
$L^{(\epsilon)}_\lambda$ on $\h$ generating the coupled time evolution in the
representation $\pi_\beta$, 
\begin{equation*}
\pi_\beta(\alpha_{t,\lambda}^{(\epsilon)}(A)) =e^{itL_\lambda^{(\epsilon)}}
\pi_\beta (A) e^{-it L_\lambda^{(\epsilon)}},
\end{equation*}
for all $A\in\frak A$ and $t\in\r$. We will show that 
\begin{equation*}
{\rm s-}\lim_{\epsilon\rightarrow 0} e^{itL_\lambda^{(\epsilon)}}
=e^{itL_\lambda}
\end{equation*}
exists, for all $t$, and defines a $*$automorphism group
\begin{equation*}
\sigma_{t,\lambda}(A)=e^{itL_\lambda}Ae^{-itL_\lambda}
\end{equation*}
of the von Neumann algebra
\begin{equation*}
\m_\beta=\pi_\beta({\frak A})''\subset {\cal B}(\h).
\end{equation*}
The pair $(\m_\beta,\sigma_{t,\lambda})$ is called a $W^*$-dynamical
system. Our results concern the structure of the set of normal
($\sigma$-weakly continuous) time-translation ($\sigma_{t,\lambda}$-) invariant states on 
$\m_\beta$. \\
\indent
The general theory of von Neumann algebras shows that there is a
one-to-one correspondence between normal $\sigma_{t,\lambda}$-invariant states
on $\m_\beta$ and normalized vectors in the set
\begin{equation}
{\cal P}\cap \ker L_\lambda,
\label{pcapker}
\end{equation}
where $\cal P$ is a certain cone in $\h$, the so-called natural cone associated to
$(\m_\beta,\Omega^{\rm ref})$, provided we choose the thermal Hamiltonian
(Liouvillian) $L_\lambda$ in such a way that the unitary one-parameter group
$\{ e^{itL_\lambda}\ |\ t\in\r\}$ leaves $\cal P$ invariant. Let $\cal M$ be a labelling of the eigenvalues
of $H_p$, including multiplicities. An element $m\in\cal M$ is called a mode
and the corresponding eigenvalue of $H_p$ is denoted by $E(m)$. We will see
that 
\begin{equation}
{\cal P}\cap \ker L_0 ={\cal P}\cap \ {\rm span}\Big\{
\varphi_m\otimes\varphi_n\otimes\Omega \ \big| \ m,n\in{\cal M},
E(m)=E(n)\Big\}, 
\label{pcapkerzero}
\end{equation}
where $\varphi_m$ is the eigenvector of $H_p$ corresponding to the mode $m$,
and $\Omega$ is the vector representative of $\omega_\beta^f$. Our main
result, Theorem \ref{regintthm}, shows that \fer{pcapker} is the subset of
\fer{pcapkerzero} with $m,n$ ranging over those modes which are not coupled to
the field. \\
\indent
While this result holds for a specific class of potentials (see \fer{2}), we
prove in Theorem \ref{nobifthm} a result which holds for a very general class
of potentials:  For any $\sigma_{t,0}$-invariant normal state $\omega^0$ and any
$\sigma_{t,\lambda}$-invariant normal state $\omega^\lambda$ on $\m_\beta$ we
prove that  $\|\omega^0-\omega^\lambda\|\geq k>0$, provided $\lambda\neq 0$ is small
enough, for a constant $k$ independent of $\lambda$. Here $\|\cdot\|$ denotes
the norm on the space of linear functionals on $\m_\beta$. Theorem \ref{nobifthm} is
proven for bounded potentials $v$ such that  $-\Delta +v$ has only finitely
many eigenvalues below the threshold of the continuous
spectrum, all of which are coupled to the field. Alternatively, we could relax
this finiteness condition but couple only finitely many modes to the field. \\

\setcounter{equation}{0}
\section{Definition of the model and main results}
In Section \ref{defmodsect} we introduce the model and show in which way it
defines a
$W^*$-dynamical system $(\m_\beta,\sigma_{t,\lambda})$. Our main results are
presented in Section \ref{mainresultsection}.

\subsection{Definition of the model}
\label{defmodsect}
Starting with an algebra $\frak A$ describing the joint system atom-field
and a (regularized)
dynamics $\alphaepsilon$ on it, we introduce a reference state $\oref$,
describing a bound state of the atom and black-body radiation at inverse
temperature $\beta$. We then 
consider the induced (regularized)
dynamics $\sigmaepsilon$ on $\pi_\beta(\frak A)$, where $(\h,\pi_\beta,\Oref)$
denotes the GNS representation corresponding to $({\frak A},\oref)$.  As 
$\epsilon\rightarrow 0$, $\sigmaepsilon$ tends 
to a $*$automorphism group, $\sigma_{t,\lambda}$, of the von Neumann algebra
$\m_\beta$, defined as the weak closure of $\pi_\beta(\frak A)$ in ${\cal
  B}(\h)$. We determine the generator, $L_\lambda$, of the unitary group, $e^{itL_\lambda}$, on
$\h$ implementing $\sigma_{t,\lambda}$; $L_\lambda$ is called a 
{\it Liouvillian}. We explain the relation between
eigenvalues of $L_\lambda$ and invariant normal states on $\m_\beta$.

\subsubsection{Kinematical algebra $\frak A$, and regularized dynamics
  $\alphaepsilon$} 
\label{subs2.1.1}

We consider a system consisting of a
quantum mechanical particle (an electron in the potential of a static nucleus)
interacting with a quantized field. \\
\indent 
Pure states of the particle system are described by unit vectors in the
Hilbert space $\h_p=L^2(\r^3,d^3x)$, their dynamics is determined by the
Schr\"o\-dinger equation with Hamiltonian
\begin{equation}
H_p=-\Delta+v,
\label{1}
\end{equation}
where the potential $v$ is bounded and satisfies one of the following two
conditions. 
\begin{itemize}
\item [$\bullet$] {\it Condition $\rm C_A$.\ } The potential $v$ is s.t. the
  spectrum of $H_p$ 
consists of a finite number $d$ of eigenvalues (counting multiplicity) lying
below the continuous spectrum which covers $[0,\infty)$. We set
$E_0:=\inf\sigma(H_p)<0$.
\item[$\bullet$] {\it Condition $\rm C_B$.\ } The potential $v$ is given by
\begin{equation}
v(x)=-\frac{\rho(|x|)}{|x|^{1+\mu}},\ \ \ \ -1<\mu\leq 1,
\label{2}
\end{equation} 
where $\rho(|x|)$ is a smooth, non-negative function that has a zero of
order $1+\mu$ at the
origin,  and increases to a constant value $\rho$ as
$|x|\rightarrow\infty$, in such a way that $v$ is smooth, and 
\begin{equation}
(x\cdot\nabla)^jv
\label{3}
\end{equation}
are bounded, for $j=0,\ldots,3$. Notice that the eigenvalues of $H_p$ are all
negative and can accumulate only at the threshold $0$.
\end{itemize}
{\it Remark.\ } In Condition $\rm C_B$ we admit potentials  such that 
$H_p$ has infinitely many eigenvalues 
below zero, but  couple only finitely many of them to the field, as we explain
below. \\

The field is a scalar massless free bosonic field. (It would be more
interesting, physically, to consider the quantized electromagnetic field. Our
methods can be applied to the resulting model at the price of slightly more
complicated notations.) The scalar free field is 
conveniently described in terms of a ``time-averaged'' Weyl algebra, ${\frak
  A}_f$, which 
is the $C^*$-algebra (of ``observables''), defined as follows. Let $\w$ be the
Weyl algebra over the Hilbert space
\begin{equation}
L_0^2=L^2(\r^3,d^3k)\cap L^2(\r^3,|k|^{-1}d^3k),
\label{L_0^2}
\end{equation}
i.e., $\w$ is the
$C^*$-algebra generated by Weyl operators, $W(f)$, $f\in L^2_0$,
satisfying the Weyl relations 
\begin{equation}
W(f)W(g)=e^{-i\IM \scalprod{f}{g}}W(g)W(f). 
\label{n39}
\end{equation}
The free field dynamics on $\w$ is given by the $*$automorphism group
\begin{equation}
W(f)\mapsto \alpha_t^\w(W(f))=W(e^{i\omega t}f),
\label{n39.1}
\end{equation}
where $\omega(k)=|k|$ is the energy of a single boson. For functions $f\in
L_0^2$, the expectation functional  
\begin{equation}
f\mapsto
e^{ -\frac{1}{4}\scalprod{f}{\left(1
      +\frac{2}{e^{\beta\omega}-1}\right)f}}
\label{sharp}
\end{equation}
is well defined and determines a $(\beta,\alpha_t^\w)$-KMS state on $\w$. 
It is well known that the 
$*$automorphism group $\alpha_t^\w$ of $\w$ is not norm-continuous (i.e.,
$\r\ni t\mapsto\alpha_t^\w(W(f))$ is not continuous in the norm of $\w$).
The time-averaged $C^*$-algebra ${\frak A}_f$ is generated by elements of the
form  
\begin{equation}
a(h)=\int_\r ds \ h(s)\alpha^\w_s(a),
\label{n40}
\end{equation}
where $a\in\w$ and $h:\r\rightarrow \cx$ are functions whose Fourier transforms
satisfy $\widehat h\in C_0^\infty$ (this is a convenient class of functions
which allows us to define KMS states on ${\frak A}_f$, see [FM]). The free field dynamics on ${\frak A}_f$
is defined by 
\begin{equation}
\alpha_t^f(a(h))=\int_\r ds\ h(s-t)\alpha_s^\w(a)=: a(h_t).
\label{n41}
\end{equation}
It is a norm-continuous $*$-automorphism group on ${\frak A}_f$. We refer to
[FM] for more details on the construction and the properties of ${\frak
  A}_f$. \\

The joint system describing the particle and the field is described in terms
of the $C^*$-algebra
\begin{equation}
{\frak A}={\frak A}_p\otimes {\frak A}_f,
\label{n42}
\end{equation}
where ${\frak A}_p={\cal B}(\h_p)$ is the von Neumann algebra of all bounded
operators on the Hilbert space $\h_p$. The uncoupled dynamics is given by the
$*$automorphism group 
\begin{equation}
\alpha_{t,0}=\alpha^p_t\otimes\alpha_t^f
\label{n43}
\end{equation}
of $\frak A$, where $\alpha^p_t(\cdot)=e^{itH_p}\cdot e^{-itH_p}$. In order to
define the dynamics of the interacting system in a representation independent way (i.e., as
a $*$automorphism group on ${\frak A}$), we need to introduce a
regularized interaction term. For $\epsilon\neq 0$, this term is given by
given by  
\begin{equation}
V^{(\epsilon)}_\#=\sum_\alpha G_{\alpha,\#}\otimes\frac{1}{2i\epsilon}\big\{ W(\epsilon
  g_\alpha)(h_\epsilon) -W(\epsilon g_\alpha)(h_\epsilon)^*\big\} \in\frak A,
\label{n44}
\end{equation}
where the sum is over finitely many indices $\alpha$, with 
$G_{\alpha,\#}=G_{\alpha,\#}^*\in{\cal B}(\h_p)$, $g_\alpha\in L^2_0$, for all
$\alpha$, and 
where $h_\epsilon$ is an approximation of the Dirac distribution localized at
zero. To be specific we can take
$h_\epsilon(t)=\frac{1}{\epsilon}e^{-t^2/\epsilon^2}$. The symbol
$G_{\alpha,\#}$ (and similarly $V_\#^{(\epsilon)}$) stands for either $G_\alpha$ or
$G_{\alpha,J}$, where $J$ is some cutoff determining which modes of the
particle are coupled to the field. In order to describe this more precisely,
we introduce the following terminology. Let $\cal M$ 
be the index set of the discrete ``modes'' of $H_p$, i.e., a labelling of the
eigenvalues of $H_p$ including multiplicity. Given $m\in\cal M$, $E(m)$
denotes the corresponding eigenvalue of $H_p$. An eigenvalue $E$
of $H_p$ is simple if and only if there is a unique $m\in\cal M$
s.t. $E=E(m)$. We denote the rank-one projection corresponding to the mode
$m\in\cal M$ by $p_m$.\\
\indent
Let $J_d\subset \cal M$ be a set of
finitely many discrete modes of $H_p$ and let
$J_c$ be an open interval in the continuous spectrum $\r_+$ of $H_p$ (we may
also 
take a finite union of disjoint intervals), s.t. $J_c\subset[r,R]$, for some $r,R$
satisfying $0<r<R<\infty$. The set  
\begin{equation*}
J:=J_d\cup J_c
\end{equation*}
determines the modes of the particle which are coupled to the field,
according to the interaction 
\begin{equation}
G_{\alpha,J}= \big(p_{J_d}+\mu(H_p)\big)G_\alpha\big(p_{J_d}+\mu(H_p)\big),
\label{54}
\end{equation}
where $G_\alpha$ is a bounded, selfadjoint operator on $\h_p$, and 
\begin{equation}
p_{J_d}=\sum_{m\in J_d} p_m,
\label{n1}
\end{equation}
$\mu\in C_0^\infty(J_c)$ is a smooth version of the indicator function with support in
$J_c$, and $\mu(H_p)$ is defined via the Fourier transform 
\begin{equation*}
\mu(H_p)=\int\widehat{\mu}(s) e^{isH_p}.
\end{equation*}
Clearly, $G_{\alpha,J}$ tends to $G_\alpha$, in
the strong sense as $\mu$ increases to the characteristic function of $\r_+$
(i.e. $J_c \uparrow \r_+$) and $J_d$ increases to the set of all discrete
modes of $H_p$. Thus, $V^{(\epsilon)}_J$ can be viewed as an approximation of
$V^{(\epsilon)}=V_{J=\r}^{(\epsilon)}$. \\ 

The interaction term \fer{n44} determines a $*$automorphism group
$\alpha_{t,\lambda}^{(\epsilon)}$ of $\frak A$, the coupled dynamics, via the
norm-convergent Dyson series  
\begin{eqnarray}
\alphaepsilon(A)&:=&\alpha_{t,0}(A) +\sum_{n\geq 1}(i\lambda)^n\int_0^t
dt_1\cdots\int_0^{t_{n-1}}dt_n \Big[
  \alpha_{t_n,0}(V_\#^{(\epsilon)}),\Big[\cdots\nonumber\\
&&\ \ \ \ \ \ \ \ \ \ \cdots
\Big[ \alpha_{t_1,0}(V_\#^{(\epsilon)}),
      \alpha_{t,0}(A)\Big]\cdots\Big]\Big],
\label{alphaepsilon}
\end{eqnarray}
where $A\in\frak A$, and $\lambda\in\r$ is the coupling constant. 
The multiple integral in
\fer{alphaepsilon} is understood in the product topology coming from the
strong topology of ${\cal B}(\h_p)$ and the norm topology of ${\frak A}_f$. \\
\indent
One may view $\alphaepsilon$
as a {\it regularized 
  dynamics}, in the sense that it has a limit, as $\epsilon\rightarrow 0$, in
suitably chosen representations of $\frak A$; (this is shown in [FM] and
explained below). 

 The functions $g_\alpha\in L^2_0$ are called
{\it form factors}. Using polar coordinates in $\r^3$, we often write
$g_\alpha=g_\alpha(\omega, \Sigma)$, where $(\omega,\Sigma)\in \r_+\times
S^2$. \\

We now specify two sets of assumptions on the interactions.\\

{\it Condition $A$.\ } The potential $v$ satisfies condition $\rm C_A$,  the
interaction is given by $V^{(\epsilon)}$, and the following properties hold. 
\begin{itemize}
\item[$\bullet$]  {\it Infrared and ultraviolet behaviour of the form
    factors}. For any fixed
  $\Sigma$, $g_\alpha(\cdot,\Sigma)\in C^4(\r_+)$, and there are two constants
  $0<k_1, k_2<\infty$, s.t. if $\omega< k_1$, then 
\begin{equation}
|\partial_\omega^j g_\alpha(\omega,\Sigma)|< k_2 \omega^{p-j}, \mbox{\ \ \ for
 some $p>2$},
\label{IR}
\end{equation}
uniformly in $\alpha$, $j=0,\ldots,4$ and $\Sigma\in S^2$. Similarly, there
are two constant $0<K_1, K_2<\infty$, s.t. if $\omega>K_1$, then
\begin{equation}
|\partial_\omega^j g_\alpha(\omega,\Sigma)|< K_2 \omega^{-q-j}, \mbox{\ \ \ for
 some $q>3$}.
\label{UV}
\end{equation}

\item[$\bullet$] {\it Relative bound on $[G_\alpha,H_p]$}. Define the
  commutator $[G_\alpha,H_p]$ in
  the weak sense  on
$C_0^\infty\times C_0^\infty$ by
\begin{equation*}
\scalprod{\psi}{[G_\alpha,H_p]\varphi}=\scalprod{G_\alpha\psi}{H_p\varphi}-\scalprod{H_p\psi}{G_\alpha\varphi}.
\end{equation*}
Then $[G_\alpha,H_p]$ extends to a relatively $(H_p-E_0+1)^{1/2}$-bounded
operator, i.e. there is a $k<\infty$ s.t. for any $\psi\in C_0^\infty$, 
\begin{equation}
\left\|[G_\alpha,H_p]\psi\right\|\leq k\left\| (H_p-E_0+1)^{1/2}\psi\right\|,
\label{relbnd}
\end{equation}
where $E_0=\inf\sigma(H_p)<0$. 
\item[$\bullet$] {\it The Fermi Golden Rule Condition}. We define a family of
  bounded operators on $\h_p$ by
$
F(\omega,\Sigma)=\sum_\alpha g_\alpha(\omega,\Sigma) G_\alpha
$
and let, for arbitrary $\epsilon>0$, 
\begin{equation}
T_\epsilon(\omega,E)=\int_{S^2}d\Sigma  \ F(\omega,\Sigma)\frac{
  p_c\ \epsilon}{(H_p-E-\omega)^2 +\epsilon^2} F(\omega,\Sigma)^*, 
\label{T}
\end{equation}
where $E$ is an eigenvalue of $H_p$ and $p_c$ is the projection onto the
 continuous subspace of $H_p$. Let $p(E)$ denote the
 projection onto the eigenspace corresponding to $E$. We assume that there is an
 $\epsilon_0>0$, s.t. for $0<\epsilon<\epsilon_0$, 
\begin{equation}
\int_{-E}^\infty d\omega\ \frac{\omega^2}{e^{\beta
    \omega}-1} p(E) T_\epsilon(\omega,E)p(E) \geq \gamma_E \ p(E),\
\label{FGRC}
\end{equation}
for any $E\in\sigma_p(H_p)$, where $\gamma_E$ is a strictly positive
 constant. We set
\begin{equation}
\gamma:=\min_{E\in\sigma_p(H_p)} \gamma_E>0.
\label{gamma}
\end{equation} 
\end{itemize}

{\it Remarks.\ } 
1) All requirements in Condition $A$ are independent of the regularization of
the interaction. \\
\indent
2)\ For the
physical model of an atom interacting with the radiation field, the value of
the constant $p$ in \fer{IR} is $p=-1/2$ (or
$p=1/2$ in the dipole approximation), see e.g. [BFS]. Although $p>2$ is quite
far from the physical range, we do not attempt here to optimize 
condition \fer{IR}. This will be the aim of subsequent work. Suffice it to
note that the discrete values $p=-1/2, 1/2$ are also admissible in our
analysis.\\
\indent
3)\  The operator $T_\epsilon(\omega,E)$
is just a (non-negative) number if $E$ is a simple eigenvalue. For $\epsilon$
small, it represents the 
probability that the particle makes a transition from the bound state
corresponding to the energy $E$ into a scattering state with energy $E+\omega\geq 0$ by
absorbing a photon of energy $\omega$. The probability density for a photon to
have energy $\omega$ is given by Planck's law, i.e., by
$(e^{\beta\omega}-1)^{-1}$. Hence $\gamma$ is a perturbative bound on the
probability of an 
ionization process; it depends on the inverse temperature
$\beta$ as $\gamma\sim e^{\beta E_0}$, where $E_0<0$ is the ground
state energy of $H_p$. More precisely, if we assume that, for
$0<\epsilon<\epsilon_0$,  $\omega\mapsto p(E)T_\epsilon(\omega,E)p(E)$ is
continuous, and that there is a constant $t>0$
s.t. $p(E)T_\epsilon(\omega,E)p(E)\geq t\cdot p(E)$ at $\omega=-E$, then 
one 
sees that $k\frac{e^{\beta E}}{1+\beta} \leq \gamma_E\leq ke^{\beta E}$, for
some $k$ which does not depend on $\beta$. \\

{\it Condition $B$.\ } The potential $v$ satisfies condition $\rm C_B$, the
interaction is given by $V_J^{(\epsilon)}$, and the following properties hold. 
\begin{itemize}
\item[$\bullet$] The infra-red and ultra-violet behaviour of the form factors
  is as in \fer{IR}, \fer{UV}.
\item[$\bullet$] Spatial decay of $G_\alpha$. There is a constant
  $k<\infty$ 
  s.t. 
\begin{equation}
\|\x^{n_1}G_\alpha\x^{n_2}\|\leq k,\ \ \ n_1+n_2=0,\ldots,5,
\label{dec}
\end{equation}
where we set $\x=(x^2+1)^{1/2}$, for $x\in\r^3$. Notice that this is a
condition on $G_\alpha$ not depending on the regularization.
\item[$\bullet$] The Fermi Golden Rule Condition. For all eigenvalues $E$ of
  $H_p$ s.t. $E=E(m)$ for some $m\in J_d$, let $T_\epsilon(\omega,E)$
  be defined as in \fer{T}, with $p_c$ replaced by $\mu(H_p)^2$, and 
let $p_{J_d}(E)=\sum_{{m\in J_d}\atop{ E(m)=E}}p_m$. There is an
  $\epsilon_0>0$ 
  s.t., for $0<\epsilon<\epsilon_0$,
\begin{equation}
\int_{-E}^\infty d\omega \ \frac{\omega^2}{e^{\beta E}-1}\ 
    p_{J_d}(E) T_\epsilon(\omega, E) p_{J_d}(E)\geq \gamma_E \ p_{J_d}(E),
\label{FGRC'}
\end{equation}
for some strictly positive constant $\gamma_E$. We set
\begin{equation}
\gamma:=\min \{\gamma_E\ |\ E\in\sigma_p(H_p) \mbox{\ s.t. $E=E(m)$ for
  some $m\in J_d$}\}>0.
\label{gamma'}
\end{equation} 
\end{itemize}

{\it Remarks.\ } 1) 
  $\gamma$ is exponentially small in $\beta$, as  observed in Remark 3)
  after \fer{gamma}.\\
\indent
2) The operator $T_\epsilon(\omega,E)$ is a decreasing function of
$r$, and an increasing function of $R$. Thus, we may assume without loss of
  generality that $\gamma$ is independent of $r\leq 1,R\geq 2$. \\

\subsubsection{Reference state $\oref$}
\label{postempsection}

The {\it reference state} of the system is given by the product state 
\begin{equation}
\omega^{\rm ref}=\omega^p\otimes\omega^f_\beta,
\label{state}
\end{equation}
where $\omega^p$ is a state on ${\cal B}(\h_p)$, determined by a strictly
positive density matrix $\rho_p>0$, i.e. 
\begin{equation}
\omega_p(A)={\rm tr}(\rho_p A), 
\label{densitymatrix}
\end{equation}
for any $A\in{\cal B}(\h_p)$. The state $\omega_\beta^f$ is the
 $\beta$-KMS 
 state of ${\frak A}_f$ w.r.t. the free field dynamics \fer{n41} determined by
 the expectation functional \fer{sharp}. It describes  
 {\it black body radiation} of the field at temperature $1/\beta$.\\
\indent
Let $(\h,\pi_\beta,\Omega^{\rm ref})$ be the GNS representation of $({\frak
  A},\omega^{\rm ref})$, i.e. $\h$ is a Hilbert space, $\pi_\beta$ is a $*$-morphism ${\frak
  A}\rightarrow {\cal B}(\h)$, and $\Omega^{\rm ref}$ is a vector in $\h$ s.t.
$\pi_\beta({\frak A})\Omega^{\rm ref}$ is dense in $\h$, and 
\begin{equation*}
\omega^{\rm ref}(A)=\scalprod{\Omega^{\rm ref}}{\pi_\beta(A)\Omega^{\rm ref}}, \ \ \ A\in{\frak
  A}. 
\end{equation*}
An explicit realization of the GNS representation is well known. It was first
constructed by Araki and Woods, [AW], and has been used recently by several
authors. Here we just recall the explicit formulas that are useful in the present
paper and refer to [JP], [FM] for a more detailed discussion. 

The representation Hilbert space is 
\begin{equation}
\h=\h_p\otimes\h_p\otimes{\cal F},
\label{imp7}
\end{equation}
where ${\cal F}$ is a shorthand for the Fock space
\begin{equation}
{\cal F}={\cal F}\left((L^2(\r\times S^2,\ du\times d\Sigma)\right),
\label{calf}
\end{equation}
$du$ being the Lebesgue measure on $\r$, and $d\Sigma$ the uniform measure on
$S^2$. Here ${\cal F}(X)$ denotes the Bosonic Fock space over a (normed vector)
space $X$,
\begin{equation}
{\cal F}(X):=\cx\oplus\bigoplus_{n\geq 1} \left({\cal S} X^{\otimes n}\right),
\label{fockspace}
\end{equation}
where ${\cal S}$ is the projection onto the symmetric subspace of the
tensor product. We use standard notation, e.g. $\Omega$ is the vacuum
vector, $[\psi]_n$ is the $n$-particle component of $\psi\in{\cal F}(X)$,
$\d\Gamma(A)$ is the second quantization of the operator $A$ on $X$,
$N=\d\Gamma(\bbbone)$ is the number operator. \\
\indent
The representation map $\pi_\beta:{\frak A}\rightarrow {\cal B}(\h)$ is the
product 
\begin{equation*}
\pi_\beta=\pi_p\otimes\pi^\beta_f,
\end{equation*}
where the $*$homomorphism 
$\pi_p:{\frak A}_p\rightarrow{\cal B}(\h_p\otimes\h_p)$ is given by
\begin{equation}
\pi_p(A)= A\otimes\bbbone_p.
\label{particlerep}
\end{equation}
The representation map $\pi_f^\beta:{\frak A}_f\rightarrow{\cal B}({\cal F})$
is determined by 
\begin{equation}
\pi^\beta_f (a(h))=\int_\r dt\ h(t)\  \pi^\beta_\w(\alpha_t^\w(a)),
\label{n45}
\end{equation}
where $\pi_\w^\beta:\w\rightarrow {\cal B}({\cal F})$ is a representation of
the Weyl algebra given by
\begin{equation*}
\pi_\w^\beta=\pi_{\rm Fock}\circ{\cal T}_\beta.
\end{equation*}
Here, ${\cal T}_\beta$ is the Bogoliubov transformation, mapping ${\frak
  W}(L^2_0)$ to ${\frak W}(L^2(\r\times 
S^2))$ defined by $W(f)\mapsto W(\tau_\beta f)$, with $\tau_\beta:
  L^2(\r_+\times 
  S^2)\rightarrow L^2(\r\times S^2)$ given by 
\begin{equation}
(\tau_\beta f)(u,\Sigma)=\sqrt{\frac{u}{1-e^{-\beta u}}}\left\{
\begin{array}{ll}
\sqrt{u}\ f(u,\Sigma), & u>0,\\
-\sqrt{-u}\ \overline{f}(-u,\Sigma), & u<0.
\end{array}
\right.
\label{imp9}
\end{equation}
{\it Remarks.} 1) It is easily verified that $\IM\scalprod{\tau_\beta
  f}{\tau_\beta g}_{L^2(\r\times S^2)}=\IM\scalprod{f}{g}_{L^2(\r_+\times S^2)}$, for all
  $f,g\in L_0^2$, so the CCR \fer{n39} are preserved under the map
  $\tau_\beta$. \\
\indent
2)\ In the limit $\beta\rightarrow\infty$, the r.h.s. of \fer{imp9} tends to
\begin{equation}
\left\{
\begin{array}{ll}
u\ f(u,\Sigma), & u>0,\\
0, & u<0.
\end{array}
\right.
\label{zero}
\end{equation}
Notice that $L^2(\r_+\times S^2)\oplus
L^2(\r_+\times S^2)$ is isometrically isomorphic to $L^2(\r\times S^2)$
via the map 
\begin{equation}
(f,g)\mapsto h,\ \ h(u,\Sigma)=\left\{
\begin{array}{ll}
u\ f(u,\Sigma), & u>0,\\
u\ g(-u,\Sigma), & u<0,
\end{array}
\right.
\label{imp9'}
\end{equation}
so \fer{zero} can be identified, via \fer{imp9'}, with $f\in
L^2_0$. Thus, 
${\cal T}_\beta$ reduces 
to the identity (an imbedding), $\pi_\w^\beta$ becomes the Fock
representation of ${\frak W}(L^2_0)$, as $\beta\rightarrow\infty$, and we
recover the zero temperature situation. \\
\indent
It is useful to introduce the following notation. We define unitary operators
$\What(f)$ on the Hilbert space 
\fer{imp7} by  
\begin{equation*}
\What(f)=e^{i\varphi(f)},\ \ f\in L^2(\r\times S^2),
\end{equation*}
where $\varphi(f)$ is the selfadjoint operator on $\cal F$ given by
\begin{equation}
\varphi(f)= \frac{a^*(f)+a(f)}{\sqrt{2}},
\label{fieldop}
\end{equation}
and $a^*(f)$, $a(f)$ are the creation- and annihilation operators on $\cal F$,
smeared out with $f$. One easily verifies that
\begin{equation*} 
\pi_\w^\beta(W(f))=\What(\tau_\beta f).   
\end{equation*}
\indent
The cyclic GNS vector is given by 
\begin{equation*}
\Oref=\Omega_p\otimes\Omega, 
\end{equation*}
where $\Omega$ is the vacuum in $\cal F$, and 
\begin{equation}
\Omega_p=\sum_{n\geq 0} k_n\varphi_n\otimes\cc_p\varphi_n\in
\h_p\otimes\h_p.
\label{omegapart}
\end{equation}
Here, $\{k^2_n\}_{n=0}^\infty$ is the spectrum of $\rho_p$,
$\{\varphi_n\}$ is an orthogonal basis of eigenvectors of $\rho_p$, and
$\cc_p$ is an antilinear involution on $\h_p$. The operator $\cc_p$ comes from 
the identification of $l^2(\h_p)$ (Hilbert-Schmitt operators on $\h_p$) with
$\h_p\otimes\h_p$, via $|\varphi\rangle\langle\psi|\mapsto
\varphi\otimes\cc_p\psi$. We fix a convenient choice for $\cc_p$. It is the
antilinear involution on $\h_p$ corresponding to complex conjugation of components of vectors in the basis in which the
Hamiltonian $H_p$ is diagonal (i.e. it is the the time reversal
operator). Then 
$
\cc_p H_p \cc_p=H_p.
$

\subsubsection{$W^*$-dynamical system $(\m_\beta,\sigma_{t,\lambda})$} 

Let $\m_\beta$ be the von Neumann algebra obtained by taking the weak closure
of $\pi_\beta({\frak A})$ in ${\cal B}(\h)$,
\begin{equation}
\m_\beta={\cal B}(\h_p)\otimes\bbbone_p\otimes\pi^\beta_f({\frak A}_f)''
\subset {\cal B}(\h).
\label{vna}
\end{equation}
Since the density matrix $\rho_p$ is strictly positive, $\Omega_p$ is
cyclic and 
separating for the von Neumann algebra $\pi_p({\frak A}_p)''={\cal
  B}(\h_p)\otimes\bbbone_p$. Similarly, $\Omega$ is cyclic and separating for
$\pi_f^\beta({\frak A}_f)''$, since it is the GNS vector of a KMS state (see
e.g. [BRII]). Consequently, $\Oref$ is cyclic and separating for
$\m_\beta$. Let $J$ be the modular conjugation operator associated to
$(\m_\beta,\Oref)$. It is given by
\begin{equation}
J=J_p\otimes J_f,
\label{J}
\end{equation}
where, for $\varphi,\psi\in\h_p$,
$
J_p\left(\varphi\otimes {\cal C}_p\psi\right)=\psi\otimes {\cal C}_p\varphi,
$ 
and, for $\psi=\{[\psi]_n\}_{n\geq 0}\in{\cal F}$,
\begin{equation*}
[J_f\psi]_n(u_1,\ldots,u_n)= \overline{[\psi]_n(-u_1,\ldots,-u_n)},\ \
\mbox{for $n\geq 1$},
\end{equation*}
and $[J_f\psi]_0=\overline{[J_f\psi]_0}\in \cx$. 
Clearly, $J\Oref=\Oref$, and one verifies that
\begin{eqnarray}
J_p\pi_p(A)J_p &=&\bbbone_p\otimes \cc_p A\cc_p,\label{jparticle}\\
J_f\pi_\w^\beta(W(f))J_f&=& \What(-e^{-\beta
  u/2}\tau_\beta(f))=\What(e^{-\beta 
  u/2}\tau_\beta(f))^*.
\label{jfield}
\end{eqnarray}
It is not difficult to see ([FM]) that  
\begin{equation}
 \sigma_{t,0}(\pi_\beta(A)):= \pi_\beta(\alpha_{t,0}(A))= e^{itL_0}\pi_\beta
  (A) e^{-itL_0},
\label{implement}
\end{equation}
for all $A\in\frak A$, where $L_0$ is the selfadjoint operator on $\h$, given
by   
\begin{equation}
L_0=H_p\otimes\bbbone_p-\bbbone_p\otimes H_p +\d\Gamma(u),
\label{L_0}
\end{equation}
commonly called the (non-interacting, standard) Liouvillian. One easily sees 
that $\alphaepsilon$ is unitarily implemented in the representation
$\pi_\beta$ as 
\begin{equation*}
\pi_\beta(\alphaepsilon(A))=
e^{itL_\lambda^{(\epsilon)}} \pi_\beta(A) 
e^{-itL_\lambda^{(\epsilon)}}=:\sigmaepsilon(\pi_\beta(A)),
\end{equation*}
where the {\it regularized Liouvillian} $L_\lambda^{(\epsilon)}$ is given by
\begin{equation*}
L_\lambda^{(\epsilon)}= L_0 +\lambda\pi_\beta(V_\#^{(\epsilon)}) -\lambda
J\pi_\beta(V_\#^{(\epsilon)}) J.
\end{equation*}
An application of the Glimm-Jaffe-Nelson Theorem (Theorem \ref{nelsonthm})
  shows that $L_\lambda^{(\epsilon)}$ is essentially selfadjoint on  
\begin{equation}
\dom=C_0^\infty\otimes C_0^\infty\otimes ({\cal F}(C_0^\infty(\r\times
S^2))\cap {\cal F}_0)\subset \h, 
\label{thedom}
\end{equation}
where ${\cal F}_0$ is the finite-particle subspace (see [FM]). Moreover, from
the theorem on invariance of domains, Theorem \ref{IDthm}, and the Duhamel
formula, one easily sees that
\begin{equation}
\lim_{\epsilon\rightarrow 0} e^{itL_\lambda^{(\epsilon)}} =e^{it L_\lambda},
\label{n46}
\end{equation}
in the strong sense on $\h$, where the Liouvillian $L_\lambda$ is given by
\begin{eqnarray}
L_\lambda&=&L_0+\lambda I,
\label{liouvillian}\\
I&=& \sum_\alpha
G_{\alpha,\#}\otimes\bbbone_p\otimes\varphi(\tau_\beta(g_\alpha)) 
-\bbbone_p\otimes \cc_pG_{\alpha,\#}\cc_p\otimes\varphi (e^{-\beta
  u/2}\tau_\beta(g_\alpha)).\nonumber
\end{eqnarray}
The operator $L_\lambda$ is essentially selfadjoint on the domain $\dom$ defined in
\fer{thedom} and 
defines a $*$automorphism group on $\m_\beta$ given by
\begin{equation}
\sigma_{t,\lambda}(A)=e^{itL_\lambda} Ae^{-itL_\lambda},\ \ \ A\in\m_\beta.
\label{dyn}
\end{equation}
An important property of $L_\lambda$ is that 
\begin{equation}
e^{it L_\lambda}J=Je^{it L_\lambda}, \mbox{\ for all $\lambda\in\r$}.
\label{n47}
\end{equation}

\subsubsection{Characterization of the $\sigma_{t,\lambda}$-invariant normal
  states on $\m_\beta$} 
\label{achsection}

A state $\omega$ on a von Neumann algebra $\m\subset{\cal B}(\h)$ is called
normal 
iff it is given by a density matrix $\rho\in{\cal B}(\h)$, i.e. 
$\omega(A)={\rm tr\,}\rho A$, $A\in\m$. If $\tau_t$ is a group of
homomorphisms of $\m$
the state is called $\tau_t$-invariant iff $\omega\circ\tau_t=\omega$ for all
$t\in\r$. In order to characterize the
$\sigma_{t,\lambda}$-invariant 
normal states on $\m_\beta$ it is useful to introduce the natural cone ${\cal
  P}$ associated to $(\m_\beta,\Oref)$, which is defined by  
\begin{equation}
{\cal P}=\overline{\{ AJA\Oref\ |\ A\in\m_\beta\}}\subset\h,
\label{natcone}
\end{equation}
where the bar denotes the closure in the norm of $\h$. 
The following properties of the natural cone are the contents of the {\it
  Araki-Connes-Haagerup} theorem, a deep result in the
theory of von Neumann algebras (see e.g. [BRI]).\\
\indent
Given any normal state $\omega$ on $\m_\beta$, there is a unique
vector $\xi\in{\cal P}$ s.t. $\omega(A)=\scalprod{\xi}{A\xi}$, for all
$A\in\m_\beta$. Moreover, if $\omega_1$ and $\omega_2$ are normal states on
$\m_\beta$ with corresponding vectors $\xi_1$, $\xi_2$ in $\cal P$ then 
\begin{equation}
\|\xi_1-\xi_2\|^2\leq \|\omega_1-\omega_2\| \leq \|\xi_1-\xi_2\|\
\|\xi_1+\xi_2\|.
\label{achbound}
\end{equation}
The norm of a state $\omega$ of $\m_\beta$ is given by
$\|\omega\|=\sup_{A\in\m_\beta} |\omega(A)|/\|A\|$.\\ 
\indent
 It is not difficult to see
  that \fer{n47} implies that $e^{itL_\lambda}{\cal P}={\cal P}$, for all
  $\lambda\in\r$ and $t\in\r$. From the uniqueness of the vector
  representative in the natural cone it follows that the
  $\sigma_{t,\lambda}$-invariant normal states are in one-to-one
  correspondence with the unit vectors in the set ${\cal P}\cap \ker
  L_\lambda$, which, for $\lambda=0$, is given by  
\begin{equation}
{\cal P}\cap \ker L_0 ={\cal P}\cap {\rm \ span}\Big\{
  \varphi_m\otimes\varphi_n\otimes\Omega \ |\ m,n\in{\cal M}, E(m)=E(n)\Big\}.
\label{i1}
\end{equation}
We will show in Theorem \ref{regintthm} that, for $\lambda\neq 0$, the
$\sigma_{t,\lambda}$-invariant normal  states are given by the
subset of \fer{i1} determined by the modes $m,n\in{\cal M}\backslash J_d$
that do not interact with the field.

\subsubsection{A quick-reference list}
\label{qrl}
For the convenience of the reader and for future reference, we collect the definitions of some
important operators in a list. Generally, if $\pi$ is a projection then we set 
$\overline\pi =\bbbone-\pi$.\\
\begin{equation*}
\begin{array}{cl}
p_d & \mbox{projection onto the discrete subspace of $H_p$}\\
p_c & \mbox{projection onto the continuous subspace of $H_p$}\\
p_m & \mbox{one-dimensional projection onto the mode $m\in{\cal M}$}\\
p_{J_d} & =\sum_{m\in J_{d}} p_m\\
p_{J_c} & \mbox{spectral projection of $H_p$ onto the interval $J_c$}\\
p & =p_{J_d}+p_{J_c}\\
P& =p\otimes p\otimes\bbbone_f\  \mbox{is a projection on $\h_p\otimes
  \h_p\otimes {\cal F}$}\\
P^l & = p\otimes \overline{p}\otimes \bbbone_f \\
P^r & =\overline{p}\otimes p\otimes \bbbone_f\\
P^0& =\overline{p}\otimes \overline{p}\otimes\bbbone_f\\
P_0 & \mbox{projection onto $\ker L_p$}\\
\Pi& =P_0\otimes P_\Omega \mbox{ is the projection onto $\ker L_0$}
\end{array}
\end{equation*}

\subsection{Main results}

\label{mainresultsection}

Our main results concern the dynamical system $({\frak
  M}_\beta,\sigma_{t,\lambda})$, where we 
have defined the von Neumann algebra ${\frak M}_\beta$ in \fer{vna},
and where $\sigma_{t,\lambda}$ is the $*$automor\-phism group \fer{dyn} of
  ${\frak 
  M}_\beta$ generated by the Liouvillian \fer{liouvillian}.  \\
\indent
The following theorem describes some properties of eigenvectors of $L_\lambda$
which, as we have seen in Subsection \ref{achsection}, play an important role
  in the characterization of invariant normal 
  states. 

\begin{theorem} {\bf (Bounds on eigenvectors).}
\label{nobifthm'}
\begin{itemize}
\item[1)] Assume that either Condition A or Condition B of Section
  \ref{subs2.1.1} 
  holds. Let 
  $N=\d\Gamma(\bbbone)$ denote the number operator on Fock space 
  ${\cal F}(L^2(\r\times S^2))$. Any eigenvector $\psi$ of $L_\lambda$
  satisfies $\psi\in\dom(N^{1/2})$, for any $\lambda\in\r$, and there is a
  constant $k<\infty$ s.t. 
\begin{equation}
\|N^{1/2}\psi\|\leq k|\lambda| \, \|\psi\|.
\label{regbnd}
\end{equation}
The constant $k$
satisfies $k<k'(1+1/\beta)$, where $k'$ depends on the interaction, but not on
$\beta$. 
\item[2)] Assume Condition A. Given any $0<\beta<\infty$, there are constants
  $\lambda_0(\beta)>0$,  $k(\beta)>0$, s.t. if
  $0<|\lambda|<\lambda_0(\beta)$, and if $\psi$ is an eigenvector of
  $L_\lambda$, then  
\begin{equation}
\|\Pbar_0\otimes P_\Omega\psi\|\geq k(\beta)\|\psi\|,
\label{regbnd'}
\end{equation}
where $P_0$ is the projection onto the zero eigenspace of $L_p$,
$\Pbar_0=\bbbone-P_0$, and $P_\Omega$ is the projection onto the
vacuum sector in ${\cal F}$. We have $\lambda_0(\beta)\geq k\gamma$  (see
\fer{gamma} and remark 2) thereafter) and $k(\beta)\geq k\gamma^2$, for some
$k$ independent of $\beta$ and $\lambda$ (i.e., both constants decay 
exponentially in $\beta$, for large $\beta$).
\end{itemize}
\end{theorem}
The proof of Theorem \ref{nobifthm'} is given in Section \ref{proof1section}.
Here we show that the bounds \fer{regbnd} and \fer{regbnd'} imply 
that bifurcations of stationary states for the interacting dynamics generated
by $L_\lambda$ from any stationary states for
$\lambda=0$ cannot occur. \\
\indent
Let $\Pi=P_0\otimes P_\Omega$ denote the projection onto the zero eigenspace
of $L_0$ and set $\Pibar:=\bbbone-\Pi$. Assume that $\psi$ is an eigenvector
of $L_\lambda$, for some $0<|\lambda|<\lambda_0$. Using the decomposition
$\Pibar=\Pbar_0\otimes P_\Omega +\Pbar_\Omega$ and \fer{regbnd},
\fer{regbnd'}, we have that  
\begin{equation*}
\left\|\Pibar\psi\right\|\geq \left\| \Pbar_0\otimes P_\Omega\psi\right\|
-\left\| \Pbar_\Omega \psi\right\| \geq (k(\beta) -k|\lambda|) \|\psi\|.
\end{equation*}
Let $\psi_0$ be an arbitrary element of $\ker L_0$. Then
\begin{equation*}
\|\psi_0-\psi\|\geq\left\|\Pibar\psi\right\| -\|\psi_0-\Pi\psi\|\geq \left\|
  \Pibar \psi\right\| -\|\psi_0-\psi\|,
\end{equation*}
so
\begin{equation}
\|\psi_0-\psi\|\geq\frac{1}{2} (k(\beta)-k |\lambda|) \|\psi\|.
\label{1.1}
\end{equation}
This shows that for $0<|\lambda|<\min(\lambda_0,
\frac{1}{2}\frac{k(\beta)}{k})$, the distance between any 
eigenvector of $L_\lambda$ and any eigenvector of $L_0$ is greater than
$k(\beta)/4$. Combining \fer{1.1} with \fer{achbound} yields the
following result.

\begin{theorem} {\bf (No bifurcation).}
\label{nobifthm}
\ Assume that the Condition $A$ of Section \ref{subs2.1.1} holds, and that
$0<|\lambda|<\min(\lambda_0, \frac{1}{2}\frac{k(\beta)}{k})$, where
$\lambda_0$, $k(\beta)$, $k$ are the constants in Theorem \ref{nobifthm'}, 2). 
For any normal $\sigma_{t,0}$-invariant state $\omega^0$ on ${\frak M}_\beta$
and any normal $\sigma_{t,\lambda}$-invariant state $\omega^\lambda$ on ${\frak
  M}_\beta$, 
\begin{equation}
\left\|\omega^0-\omega^\lambda\right\|\geq k(\beta)^2/16.
\label{posdiff}
\end{equation}
\end{theorem}
\ \ \ \ 
Our next result shows that the modes of the particle which are coupled to the
field do not give rise to invariant states; (compare with Section \ref{achsection}).
\begin{theorem}{\bf (Instability of normal invariant states).\ }
\label{regintthm} Assume that Condition $B$ of Section \ref{subs2.1.1}
holds. Given any $0<\beta<\infty$ and any 
$r>0$, there is a $\lambda_0(\beta,r)>0$ s.t. for
  $0<|\lambda|\leq\lambda_0(\beta,r)$, the $\sigma_{t,\lambda}$-invariant
  normal states on $\m_\beta$ are in one-to-one correspondence with the unit
  vectors in the set
\begin{equation}
{\cal P}\cap {\rm \ span} \Big\{ \varphi_m\otimes\varphi_n\otimes \Omega\ |\
m,n\in {\cal M}\backslash J_d, \ E(m)=E(n)\Big\}.
\label{res1}
\end{equation}
We have $\lambda_0(\beta,r)\geq
  k\gamma^2r$, for some $k$ which is independent of $\beta,r$ and where
  $\gamma$ is given in \fer{gamma'}.
\end{theorem}

\setcounter{equation}{0}
\section{Virial theorems and the positive commutator method}
Our proofs of Theorems \ref{nobifthm'} and \ref{regintthm} are based on the
positive commutator method, which we explain in Section
\ref{pcmethodsection}. In the next section we describe an essential ingredient
of this method, the virial theorem. 

\subsection{Two abstract virial theorems}
\label{absvthm}

Let $\h$ be a Hilbert space, $\dom\subset\h$ a core for a selfadjoint
operator $Y\geq\bbbone$, and $X$ a symmetric operator on 
$\dom$. We say the triple $(X,Y,\dom)$ satisfies the {\it GJN
  (Glimm-Jaffe-Nelson) 
  Condition}, or that $(X,Y,\dom)$ is a {\it GJN-triple}, if there is a
constant $k<\infty$, s.t. for all $\psi\in\dom$: 
\begin{eqnarray}
\|X\psi\|&\leq& k\|Y\psi\| \label{nc1}\\
\pm i\left\{\scalprod{X\psi}{Y\psi}-\scalprod{Y\psi}{X\psi}\right\}&\leq&
k\scalprod{\psi}{Y\psi}.
\label{nc2}
\end{eqnarray}
Notice that if $(X_1,Y,\dom)$ and $(X_2,Y,\dom)$ are GJN triples, then so is
$(X_1+X_2,Y,\dom)$. Since $Y\geq\bbbone$, inequality \fer{nc1} is equivalent
to  
\begin{equation*}
\| X\psi\|\leq k_1\|Y\psi\|+k_2\|\psi\|,
\end{equation*}
for some $k_1, k_2<\infty$. For a more detailed exposition of the following
results (and proofs of Theorems \ref{virialthm} and \ref{virialthm'}) we
refer to [FM].

\begin{theorem}{\bf (GJN commutator theorem) }
\label{nelsonthm}
\stepcounter{proposition}
If $(X,Y,\dom)$ satisfies
  the GJN Condition, then $X$ determines a selfadjoint operator (again
  denoted by $X$), s.t. $\dom(X)\supset\dom(Y)$. Moreover, $X$ is essentially
  selfadjoint on any core for $Y$, and \fer{nc1} is valid for all
  $\psi\in\dom(Y)$.
\end{theorem}

Based on the GJN commutator theorem, we next
describe the setting for a general {\it virial theorem}. Suppose one is given
a selfadjoint operator $\Lambda\geq\bbbone$ with core $\dom\subset\h$, and 
 operators  $L, A, N, D, C_n$, $n=0,1,2,3$, all symmetric on $\dom$, and
 satisfying  
\begin{eqnarray}
\scalprod{\varphi}{D\psi}&=&i\left\{
  \scalprod{L\varphi}{N\psi}-\scalprod{N\varphi}{L\psi}\right\} \label{44}\\
C_0&=& L\nonumber\\
\scalprod{\varphi}{C_n\psi}&=&i\left\{\scalprod{C_{n-1}\varphi}{A\psi}-\scalprod{A\varphi}{C_{n-1}\psi}\right\},\
  \ n=1,2,3,
\label{45}
\end{eqnarray}
where $\varphi, \psi\in\dom$. We assume that
\begin{itemize}
\item[$\bullet$]  $(X,\Lambda,\dom)$ satisfies the GJN Condition, for
$X=L,N,D,C_n$. Consequently, all these operators determine selfadjoint
operators, which we denote by the same letters.
\item[$\bullet$] $A$ is selfadjoint, $\dom\subset\dom(A)$, and 
$e^{itA}$ leaves $\dom(\Lambda)$
invariant.
\end{itemize}
{\it Remarks.\ } 
1) From the invariance condition $e^{itA}\dom(\Lambda)\subset \dom(\Lambda)$, it follows that for some $0\leq k,k'<\infty$, and all $\psi\in\dom(\Lambda)$,
\begin{equation}
\|\Lambda e^{itA}\psi\|\leq ke^{k'|t|}\|\Lambda\psi\|.
\label{alambda}
\end{equation}
A proof of this can be found in [ABG], Propositions 3.2.2 and 3.2.5.\\
2)\ Condition \fer{nc1} is phrased equivalently as ``$X\leq
kY$, in the sense of Kato on $\dom$''. \\
3)\ One can show that if $(A,\Lambda,\dom)$ satisfies conditions \fer{nc1},
\fer{nc2}, then the above assumption on $A$ holds; see Theorem \ref{IDthm}.

\begin{theorem}{\bf ($1^{\rm st}$ virial theorem) }
\label{virialthm}
Assume $N$ and $e^{itA}$ commute, for all $t\in\r$, in the strong sense on
  $\dom$, and that
\begin{eqnarray}
 D&\leq& kN^{1/2},\label{41}\\
 C_1&\leq& kN^{p},\ \ \mbox{\ for some $0\leq p <\infty$,}\label{40}\\
C_3&\leq& kN^{1/2}\label{40'}
\end{eqnarray}
in the sense of Kato on $\dom$, for some $k<\infty$. 
Then, if $\psi\in\dom(L)$ is an eigenvector of $L$,
there is a family of approximating eigenvectors
$\{\psi_\alpha\}\subset\dom(L)\cap\dom(C_1)$, $\alpha>0$, such that
$\psi_\alpha\rightarrow\psi$ (in $\h$) as $\alpha\rightarrow 0$, and 
\begin{equation}
\lim_{\alpha\rightarrow 0}\scalprod{\psi_{\alpha}}{C_1\psi_{\alpha}}=0.
\label{42}
\end{equation}
\end{theorem}

{\it Remarks.\ }\ 1) It is not necessary that $N$ and $e^{itA}$ commute for
the result to hold, but this will be the case in all our applications (see
[FM] for the case where $N$ and $A$ do not commute).\\
2) In a heuristic way, we understand $C_1$ as the commutator
$i[L,A]=i(LA-AL)$, and \fer{42} as $\scalprod{\psi}{i[L,A]\psi}=0$, which is
the usual statement of the virial theorem; see e.g. [ABG], and
[GG] for a comparison (and correction) of virial theorems encountered in the
literature. \\ 

The Virial Theorem is still valid if we add to the operator $A$ a bounded
perturbation $A_0$ leaving the domain of $L$ invariant.
\begin{theorem}{\bf ($2^{\rm nd}$ virial theorem) }
\label{virialthm'}
 Suppose that we are in the situation of Theorem \ref{virialthm}, and that
  $A_0$ is a bounded operator on $\h$, s.t. $\ran 
  A_0\subseteq\dom(L)\cap\ran P(N\leq n_0)$, for some $n_0<\infty$. The
  commutator 
  $i[L,A_0]=i(LA_0-A_0L)$ is well defined in the strong sense on $\dom(L)$.
  For the same family of approximating eigenvectors as in the
  previous theorem, we have that
\begin{equation}
\lim_{\alpha\rightarrow 0}\scalprod{\psi_{\alpha}}{(C_1+i[L,A_0])\psi_{\alpha}}=0.
\label{43}
\end{equation}
\end{theorem}

\subsection{Outline of the proofs of Theorems \ref{nobifthm'} and
  \ref{regintthm}; the positive commutator method}
\label{pcmethodsection}

The positive commutator method gives a conceptually easy proof of the absence
of point spectrum of $L$. We outline a version that is adapted to the proofs
of Theorems \ref{nobifthm'} and \ref{regintthm}. The full proofs are given in
Sections \ref{proof1section} and \ref{proof2section}.
In the present section we use the notation of Section \ref{absvthm} and write
$i[L_\lambda,A]$ for $C_1$.\\

{\it Outline of the proof of Theorem \ref{regintthm}.\ }
According to the discussion of Section \ref{achsection} we have to show that
${\cal P}\cap\ker L_\lambda$ is given by the set \fer{res1}. The Liouvillian
$L_\lambda$ is reduced by the decomposition 
\begin{equation*}
\h=\ran P^0\oplus\ran P\oplus\ran P^l\oplus\ran P^r,
\end{equation*}
where the various projections are defined in Section \ref{qrl}. It is easy to
see that ${\cal P}\cap\ran P^l={\cal P}\cap\ran P^r=\{0\}$ and that
$L_\lambda\upharpoonright \ran P^0=L_0\upharpoonright \ran P^0$. Consequently,
$
{\cal P}\cap \ker L_\lambda ={\cal P}\cap\left( \ker L_0\upharpoonright\ran
  P^0 \cup \ker L_\lambda\upharpoonright\ran P\right),
$ 
and to prove the theorem, it is enough to show that 
\begin{equation}
\ker L_\lambda\upharpoonright \ran P=\{0\}.
\label{3.2.1}
\end{equation}
We construct selfadjoint operators $\Lambda, A, A_0$ such that the conditions
of Section \ref{absvthm} are fulfilled, with $L=L_\lambda$ and
$N=\d\Gamma(\bbbone)$. The operators $A$ and $A_0$ have the properties that 
\begin{equation*}
i[L_\lambda,A]+i[L,A_0]\geq PM_0P +M_1,
\end{equation*}
where the bounded operators $M_0$ and $M_1$ satisfy
\begin{eqnarray}
PM_1P&=&0,\nonumber\\
\scalprod{\psi}{M_0\psi}&\geq&\delta\|\psi\|^2,
\label{3.2.2}
\end{eqnarray}
for any $\psi\in\ker L_\lambda\upharpoonright\ran P$, and where $\delta$ is
strictly positive. From Theorem \ref{virialthm'} we obtain
\begin{equation*}
0=\lim_\alpha\scalprod{\psi_\alpha}{\left(
    i[L_\lambda,A]+i[L_\lambda,A_0]\right)\psi_\alpha} \geq \delta\|\psi\|^2.
\end{equation*}
Because $\delta>0$, we have that $\ker L_\lambda\upharpoonright \ran
P=\{0\}$. \\
\indent
We now explain how to arrive at the key inequality \fer{3.2.2}. The form
of the non-interacting Liouvillian, $L_0$, given in \fer{L_0} suggests to
consider 
\begin{equation*}
A=\chi A_p\chi\otimes\bbbone_p\otimes\bbbone_f -\bbbone_p\otimes\chi
A_p\chi\otimes \bbbone_f +\bbbone_p\otimes\bbbone_p\otimes
\d\Gamma(i\partial_u),
\end{equation*}
where $A_p$ is the dilation generator, see \fer{ap}, and
$\d\Gamma(i\partial_u)$ is the second quantization of the translation
generator in the radial variable 
$u$ of $L^2(\r\times S^2,du\times d\Sigma)$, see \fer{calf}. Here, $\chi$ is a function of $H_p$ with support in an interval in
$\r_+$, containing $J_c$ but not $\{0\}$, and such that $\chi|_{J_c}=1$. Then
we have 
\begin{equation*}
i[L_\lambda,A]=\chi^2H_p\otimes\bbbone_p\otimes\bbbone_f +\bbbone_p\otimes
\chi^2H_p\otimes\bbbone_f +N+U+\lambda I_1,
\end{equation*}
where $N$ is the number operator, $U$ is a non-negative operator, because of
the choice of the parameter $\mu$ in the potential $v$ of equation \fer{2}, and where $I_1=i[I,A]$ is
infinitesimally small w.r.t. $N$, 
\begin{equation}
\pm\lambda I_1 \leq cN +\frac{\lambda^2}{c}k,
\label{3.2.3}
\end{equation}
for any $c>0$ and some $k<\infty$. The role of $\chi$ is to project out the
discrete modes in $J_d$. Using that $P=p\otimes p\otimes\bbbone_f$,
$p\chi^2=p_{J_c}$, and that $p_{J_c}H_p\geq rp_{J_c}$, since the interval
$J_c$ is away from the origin by a distance of at least $r$, we obtain
\begin{eqnarray*}
\lefteqn{
Pi[L_\lambda,A]P}\\
&\geq& P\left( p_{J_c} H_p\otimes\bbbone_p\otimes P_\Omega
  +\bbbone_p\otimes p_{J_c}H_p\otimes P_\Omega +\frac{1}{2}
  \Pbar_\Omega\right) P -k\lambda^2 P\\
&\geq&\min(r,1/2) P\left( p_{J_c} \otimes\bbbone_p\otimes P_\Omega
  +\bbbone_p\otimes p_{J_c}\otimes P_\Omega +\frac{1}{2}
  \Pbar_\Omega\right) P -k\lambda^2 P,
\end{eqnarray*}
where we choose $c=1/2$ in \fer{3.2.3}. The first term on the r.h.s. is
strictly positive except on the subspace $\ran p_{J_d}\otimes p_{J_d}\otimes
P_\Omega\subset\ran P$. We decompose
\begin{equation}
 p_{J_d}\otimes p_{J_d}\otimes P_\Omega =P \Pi +\sum_{{m,n\in J_d}\atop
  {E(m)\neq E(n)}} p_m\otimes p_n\otimes P_\Omega,
\label{3.2.4}
\end{equation}
where $\Pi$ is the projection onto the kernel of $L_0$. Note that $P\Pi$ is finite-dimensional. On the range of the second projection on the r.h.s. of
\fer{3.2.4}, the free Liouvillian satisfies
\begin{equation}
|L_0|\geq \min\Big\{ |E(m)-E(n)|\ \big|\ m,n\in J_d, E(m)\neq E(n)\Big\}>0.
\label{3.2.5}
\end{equation}
Let $\Delta$ be an interval around zero whose size $|\Delta|$ is smaller than
the r.h.s. of \fer{3.2.4}, and let $E_\Delta^0$ be the spectral projection of
$L_0$ onto $\Delta$. Then we have that $E_\Delta^0\  p_{J_d}\otimes p_{J_d}\otimes
P_\Omega= E_\Delta^0P\Pi=P\Pi$. Consider the decomposition
\begin{equation}
\ran PE_\Delta^0 =\ran P E_\Delta^0\Pbar\oplus\ran P\Pi.
\label{3.2.6}
\end{equation}
From the above discussion it is apparent that on the block $\ran P
E_\Delta^0\Pbar$, $i[L_\lambda,A]$ is bigger than $r-k\lambda^2\geq r/2$ (we
require $|\lambda|\leq k\sqrt r$), while the commutator is zero on the block
$\ran P\Pi$. This is where we introduce the operator $A_0$.\\
\indent
One can choose $A_0$ s.t. $P\Pi i[L_\lambda,A_0]\Pi P$ is strictly positive
provided the interaction satisfies the Fermi Golden Rule
Condition. Moreover, on $\ran PE_\Delta^0\Pibar$, $i[L_\lambda,A_0]$ is small
relative to $r$. The construction of $A_0$ has been given, in the context
of zero temperature systems, in [BFSS], and has been modified for positive
temperature systems in [M]. \\
\indent
The above discussion shows that the operator $i[L_\lambda,
A]+i[L_\lambda,A_0]$ has strictly positive diagonal blocks in the
decomposition \fer{3.2.6}. An application of the Feshbach method then shows
that 
\begin{equation}
E_\Delta^0 P\left(i[L_\lambda,A] +i[L_\lambda, A_0]\right) P E_\Delta^0=:
E_\Delta^0 P M_0 P E_\Delta^0
\label{3.2.7}
\end{equation}
is strictly positive. Since $\| E_\Delta^0\psi-\psi\|\leq k|\lambda|\ \|\psi\|$, for any
$\psi\in\ker L_\lambda$, we can pass from \fer{3.2.7} to estimate
\fer{3.2.2}.\\
\indent
In this proof, the coupling constant cannot be chosen idependently of
the inverse temperature $\beta$. This is due to the fact that the constant
$\delta$ in \fer{3.2.2} is proportional to $\gamma$, see \fer{gamma'}, which
in turn 
decays exponentially in $\beta$. We have to require that certain error terms
which depend on $\lambda$ are small w.r.t. $\gamma$, hence the
$\beta$-dependent smallness condition on $\lambda$. \\ 

{\it Outline of the proof of Theorem \ref{nobifthm'}\ }. Part 1) is an
easy consequence of the virial theorem combined with the bound
\begin{equation*}
i[L_\lambda, A]\geq \frac{1}{2} N -k\lambda^2,
\end{equation*}
for $A=\d\Gamma(i\partial_u)$. The proof of part
2) proceeds as follows. We construct $A_0$ (the same as for the proof
of Theorem \ref{nobifthm'}) s.t.
\begin{equation}
i[L_\lambda,A]+i[L_\lambda,A_0]\geq \kappa_1 \Pi -\kappa_2 \Pbar_0\otimes P_\Omega,
\label{pc3}
\end{equation}
for some $\kappa_1, \kappa_2>0$, and where $\Pi$ is the projection onto the
kernel 
of $L_0$. If $\psi$ is an eigenvector of $L_\lambda$ then by part 1) we have
that 
$\|\Pi\psi\|\geq (1-k|\lambda|)\|\psi\|-\|\Pbar_0\otimes
P_\Omega\psi\|$. Inserting this bound into  \fer{pc3} and using the virial
theorem yields the bound \fer{regbnd'}.\\ 

This outline indicates that the proofs of Theorems \ref{nobifthm'}
and \ref{regintthm} consist of two steps. First we verify that 
the virial theorems are applicable and second establish a positive
commutator 
estimate in the above sense. The latter task is carried out in Sections
\ref{proof1section} and \ref{proof2section}.

\subsection{Applications of the virial theorems}

Corresponding to the different hypotheses of Theorems \ref{nobifthm'} and
\ref{regintthm}, we 
introduce two sets of operators $\Lambda,L,A,A_0,N$, and verify, in each case,
that the virial theorems are applicable. The following objects appear in
both applications: the Hilbert space is the GNS space given in \fer{imp7}; the
dense domain $\dom$ is chosen to be 
\begin{equation}
\dom=C_0^\infty(\r^3)\otimes C_0^\infty(\r^3)\otimes\dom_f,
\label{44'}
\end{equation}
where 
\begin{equation*}
\dom_f={\cal F}\left(C_0^\infty(\r\times S^2)\right)\cap {\cal F}_0,
\end{equation*}
where the Fock space ${\cal F}$ has been defined in \fer{fockspace},  
and ${\cal F}_0$ denotes the finite-particle subspace. The
operator $L$ is the interacting Liouvillian introduced in \fer{liouvillian},
and $N=\d\Gamma(\bbbone)$ 
is the particle   
number operator in ${\cal F}\equiv {\cal F}(L^2(\r\times S^2))$. Clearly,
$X=L,N$ are 
symmetric operators on $\dom$. The operator $D$, defined in \fer{44}, 
is given by   
\begin{eqnarray}
D&=&i\lambda\sum_\alpha\big\{G_{\alpha,\#}\otimes\bbbone_p
    \otimes\left(-a^*(\tau_\beta(g_\alpha))+  
    a(\tau_\beta(g_\alpha))\right)\nonumber\\
&&-\bbbone_p\otimes {\cal C}_p G_{\alpha,\#}{\cal C}_p \otimes\left(
    -a^*(e^{-\beta u/2}\tau_\beta(g_\alpha))+a(e^{-\beta
    u/2}\tau_\beta(g_\alpha))\right)\big\}.\ \ \ \ \ \ \ \ 
\label{operatorD}
\end{eqnarray}

We define a bounded, selfadjoint operator
$A_0$ on $\h$ by 
\begin{eqnarray}
A_0&=&i\theta\lambda (\Pi
  I\repsilon^2\Pibar -\Pibar\repsilon^2I\Pi),\label{17}\\
\repsilon^2&=&(L_0^2+\epsilon^2)^{-1}.\label{18}
\end{eqnarray}
Here, $\theta$ and $\epsilon$ are positive  parameters, and $\Pi$ is the
projection 
\begin{eqnarray}
\Pi&=&P_0\otimes P_\Omega,\label{19}\\
P_0&=&P(L_p=0),\label{20}\\
\Pibar&=&\bbbone-\Pi\label{21}.
\end{eqnarray}
We also introduce the notation 
\begin{equation*}
\repsilonbar=\Pibar \repsilon.
\end{equation*}
\indent
Notice that the operator $A_0$ satisfies the conditions given in  Theorem
\ref{virialthm'} with $n_0=1$. Moreover, $[L_,A_0]=L A_0-A_0L$ extends to a bounded operator on the entire Hilbert space, and 
\begin{equation}
\|[L,A_0]\|\leq k\left(\frac{\theta
    |\lambda|}{\epsilon}+\frac{\theta\lambda^2}{\epsilon^2}\right). 
\label{a}
\end{equation}
\indent
This choice for the operator $A_0$ was initially introduced in [BFSS] for the
spectral analysis of Pauli-Fierz Hamiltonians (zero temperature systems), and
was adopted in [M] to show return to equilibrium (positive temperature
systems). The key feature of $A_0$ is that 
\begin{equation*}
i\Pi[L,A_0]\Pi=2\theta\lambda^2
\Pi I \repsilonbar^2 I\Pi
\end{equation*}
is a non-negative operator. The Fermi
Golden Rule Condition, \fer{FGRC} (or \fer{FGRC'}), says that it is a {\it
  strictly positive operator} on $\ran \Pi$. 
\begin{proposition}
\label{fgrprop}
Assume \fer{FGRC} and let $0<\epsilon<\epsilon_0$. Then
\begin{equation}
\Pi I \repsilonbar^2 I \Pi \geq \frac{1}{\epsilon}\gamma\Pi,
\label{fgrc1}
\end{equation}
where $\gamma$ is given by \fer{gamma}. 
Assuming condition \fer{FGRC'} instead of \fer{FGRC}, the same lower bound
holds (with $\gamma$ given in 
\fer{gamma'}) if we replace $\Pi$ by $\Pi P$
($P=p\otimes p$, $p=p_{J_d}+p_{J_c}$) and $I$ by the regularized interaction;
see \fer{54}.
\end{proposition}

The proof of Proposition \ref{fgrprop} is given in Section
\ref{propproofsection}.\\
Next, we define the operators $\Lambda$ and $A$ and verify the hypotheses
used in Section \ref{absvthm}.

\subsubsection{Setting for Theorem \ref{nobifthm'}}
\label{sub1}
We define
\begin{eqnarray}
\Lambda&=&\Lambda_p\otimes\bbbone_p\otimes\bbbone_f+\bbbone_p\otimes\Lambda_p\otimes\bbbone_f+\bbbone_p\otimes\bbbone_p\otimes\Lambda_f,
\label{46}\\
\Lambda_p&=& H_p-E_0+1,\label{47'}\\
\Lambda_f&=&\d\Gamma(u^2+1),
\label{48}
\end{eqnarray}
where, we recall, $E_0=\inf\sigma(H_p)<0$. 
Clearly, $\Lambda$ is essentially selfadjoint on the domain $\dom$ defined in \fer{44'}, and
$\Lambda_p\geq\bbbone$, $\Lambda_f\geq \bbbone$.
In what follows we shall
often use the standard fact that if $f\in L^2(\r\times S^2, du\times
d\Sigma)$, then $a^\#(f)$ is relatively $N^{1/2}$ bounded in the sense of
Kato. This implies immediately that $a^\#(f)$ is relatively $\Lambda_f^{1/2}$
bounded.\\ 
\indent 
We verify that $(L,\Lambda,\dom)$ is a GJN triple. The bound \fer{nc1} is
trivial 
by the above observation, and the fact that $\tau_\beta(g_\alpha)\in L^2(\r\times
S^2)$. Next, the only contribution to the commutator of $L$ with $\Lambda$
comes from the interaction, and a typical term to estimate is of the form
$[G_\alpha,H_p]\otimes\bbbone\otimes\varphi(\tau_\beta(g_\alpha))
+G_\alpha\otimes\bbbone_p\otimes[\varphi(\tau_\beta(g_\alpha)),\Lambda_f]$. Using
the bound \fer{relbnd}, we obtain for the first term
\begin{eqnarray}
\lefteqn{
\left|\scalprod{\psi}{[G_\alpha,H_p]\otimes\bbbone_p\otimes\varphi(\tau_\beta(g_\alpha))\psi}\right|}\nonumber\\
&\leq&
k \|[G_\alpha,H_p](H_p-E_0+1)^{-1/2}\|\ 
\|\Lambda_p^{1/2}\otimes\bbbone_p\psi\|\ \|\Lambda_f^{1/2}\psi\|\nonumber\\
&\leq& k\scalprod{\psi}{\Lambda\psi}.
\label{d2}
\end{eqnarray} 
Next, 
\begin{eqnarray}
\left|\scalprod{\psi}{G_{\alpha}\otimes\bbbone_p\otimes[\varphi(\tau_\beta(g_\alpha)),\Lambda_f]\psi}\right| 
&\leq& k\|(u^2+1)^{1/2}\tau_\beta(g_\alpha)\|_{L^2(\r\times S^2)}\
\|\Lambda^{1/2}\psi\|^2\nonumber\\
&\leq& k\scalprod{\psi}{\Lambda\psi},
\label{d3}
\end{eqnarray}
where we have used that
\begin{eqnarray*}
\left[a^*(\tau_\beta(g_\alpha)),\Lambda_f\right] &=& a^*\left(
  (u^2+1)\tau_\beta(g_\alpha)\right),\\
\left[a(\tau_\beta(g_\alpha)),\Lambda_f\right] &=& -a\left(
  (u^2+1)\tau_\beta(g_\alpha)\right),
\end{eqnarray*}
so that $[\varphi(\tau_\beta(g_\alpha)),\Lambda_f]$ is still $N^{1/2}$
bounded, since $\tau_\beta(g_\alpha)$ has the decay 
property \fer{UV}. The form bound \fer{nc2} follows from these observations.
In a similar way, one shows that $(D,\Lambda,\dom)$ is a GJN triple.\\

 Next, we define the operator
$A\equiv A_f$ to be the selfadjoint generator of the translation group acting
on the radial variable of elements in ${\cal F}$ by 
\begin{equation*}
[e^{itA_f}\psi]_n(u_1,\Sigma_1,\ldots,u_n,\Sigma_n)=
[\psi]_n(u_1-t,\Sigma_1,\ldots,u_n-t,\Sigma_n), \ t\in\r.
\end{equation*}
In what follows, we will often not display the angular variables
$\Sigma_1,\ldots,\Sigma_n$. We 
set $e^{itA_f}\Omega:=\Omega$. Clearly, $\dom\subset\dom(A_f)$, $\dom$ is
invariant under $e^{itA_f}$, 
hence a core for $A_f$, and $A_f$ acts on $\dom$ as 
\begin{equation}
A_f=\d\Gamma(i\partial_u).
\label{af}
\end{equation}
An easy calculation shows that, on $\dom$,
\begin{equation}
\Lambda e^{itA_f}=e^{itA_f}\left(\Lambda+\d\Gamma(2ut-t^2)\right),
\label{easycalc}
\end{equation}
so estimate \fer{alambda}, with $k'=0$, is satisfied for all
$\psi\in\dom$, hence for all $\psi \in \dom(\Lambda)$. 
For $A:=A_f$, we find that
\begin{eqnarray}
C_1&=&N+\lambda I_1, \label{50}\\
C_2&=& \lambda I_2, \label{51}\\
C_3&=& \lambda I_3,\label{52}
\end{eqnarray}
where 
\begin{eqnarray}
\lefteqn{
I_n=i^n\sum_\alpha\big\{
G_\alpha\otimes\bbbone_p\otimes\varphi((-i\partial_u)^n\tau_\beta(g_\alpha))}\nonumber\\
&&-\bbbone_p\otimes{\cal
  C}_p G_\alpha{\cal C}_p\otimes\varphi((-i\partial_u)^ne^{-\beta
  u/2}\tau_\beta(g_\alpha))\big\}.
\label{25}
\end{eqnarray}
\indent
We now show that $(C_n,\Lambda,\dom)$ are GJN triples, for $n=1,2,3$. 
The operators $I_n$ are $N^{1/2}$-bounded, since $\tau_\beta(g_\alpha)$
and $e^{-\beta u/2}\tau_\beta(g_\alpha)$ are in in the domain of the operators
$(i\partial_u)^n$, $n=1,2,3$ (see also \fer{IR}), hence \fer{nc1}
holds. Note that 
this also yields \fer{40}. Next, we need to calculate the commutators of $C_n$
with $\Lambda$. The estimates on the commutators of
$I_n$ with $\Lambda$, for $n=2,3$, are similar to the ones for $n=1$. The
latter has 
been outlined above; it requires that   
$(u^2+1)(-i\partial_u)^n\tau_\beta(g_\alpha)\in L^2(\r\times
S^2)$, which is guaranteed by conditions \fer{IR} and \fer{UV}.\\
\indent
This discussion shows that we are in the
situation described in Section \ref{absvthm}, and Theorems \ref{virialthm},
\ref{virialthm'} apply.

\subsubsection{Setting for Theorem \ref{regintthm}}
\label{sub2}
We define the operator $\Lambda$ as in \fer{46}, but where $\Lambda_p$ is now
given by 
\begin{equation}
\Lambda_p=-\Delta+x^2.
\label{47}
\end{equation}
$\Lambda$ is essentially selfadjoint on $\dom$ (see \fer{44'}), 
$\Lambda_p\geq\bbbone$, $\Lambda_f\geq \bbbone$.\\
\indent
Verifying that $(L_J,\Lambda,\dom)$ is a GJN triple is done as in
Subsection \ref{sub1},  using that $[G_{\alpha,J},\Lambda_p]$ is bounded;
see Lemma \ref{reglemma}. It is also easy to check that $(D,\Lambda,\dom)$ is
a GJN triple.\\

Next, we define an operator
$A$ differing substantially from the choice $A=A_f$ (see \fer{af}) in
Subsection \ref{sub1}: We add a (regularized) dilatation on the
particle space to $A_f$. \\ 
\indent
Let $\chi\in C_0^\infty(\r_+)$ be a smooth characteristic function of the
set  $J_c$ (with the property $\chi|_{J_c}=1$), which has compact support not
containing zero. We define 
$\chi(H_p)=\int\hat{\chi}(s)e^{isH_p}$, where $\hat{\chi}$ is the Fourier
transform of $\chi$, and we abbreviate $\chi(H_p)$ by $\chi$. Let
$A_p$ be the symmetric operator on $C_0^\infty(\r^3)$ given by
\begin{equation}
A_p=-\frac{i}{4}(x\cdot\nabla+\nabla\cdot x).
\label{ap}
\end{equation}
Notice that $(A_p,\Lambda_p,C_0^\infty(\r^3))$ is a GJN triple, so $A_p$ is
essentially selfadjoint on $C_0^\infty(\r^3)$. We denote the selfadjoint
closure again by $A_p$. \\
\indent
{\it Remark.\ } To show that $A_p$ is essentially
selfadjoint on $C_0^\infty(\r^3)$, we can also use the fact that the 
dense set $C_0^\infty(\r^3)$ is invariant under the group 
of dilatations on $L^2(\r^3,d^3x)$, hence a core for the selfadjoint generator
of this group. The generator acts on $C_0^\infty$ as in \fer{ap}. 
\begin{proposition}
\label{Aprop}
$(\chi A_p\chi,\Lambda_p,C_0^\infty(\r^3))$ is a GJN triple. In particular,
$\chi A_p\chi$ is well defined and symmetric on $C_0^\infty(\r^3)$, and it is
essentially selfadjoint on $C_0^\infty(\r^3)$. We denote the
selfadjoint closure again by $\chi A_p\chi$.
\end{proposition}
We give the proof in Section \ref{proofprop3.2section}. Let us now define the
operator 
\begin{equation}
A=\chi A_p\chi\otimes\bbbone_p-\bbbone_p\otimes\chi A_p\chi +A_f,
\label{55}
\end{equation}
which is essentially selfadjoint on $\dom$. It follows immediately
from Proposition \ref{Aprop}, Theorem \ref{IDthm}, and relation
\fer{easycalc}, that $e^{itA}$ leaves $\dom(\Lambda)$ invariant, and that the
estimate \fer{alambda} holds true. We calculate explicitly
\begin{equation}
C_n=\delta_{n,1}N +{\rm ad}^{(n)}_{\chi
  A_p\chi}(H_p)\otimes\bbbone_p +(-1)^n\bbbone_p\otimes{\rm ad}_{\chi A_p
  \chi}^{(n)}(H_p)+\lambda I_n, \label{56}
\end{equation}
for $n=1,2,3$, where we define the multiple commutators ${\rm
  ad}^{(0)}_Y(X)=X$, and for $n\geq  
1$, ${\rm ad}^{(n)}_Y(X)=i[{\rm ad}_Y^{(n-1)}(X),Y]$, in the weak sense on
$C_0^\infty(\r^3)\times C_0^\infty(\r^3)$. For 
$n=1,2,3$, we have defined 
\begin{eqnarray}
I_n&=&\sum_{k=0}^n {n \choose k}2^{-(n-k)}\sum_\alpha \left\{ {\rm ad}_{\chi
  A_p\chi}^{(n-k)}(G_\alpha)\otimes\bbbone_p\otimes\varphi\left((-i\partial_u)^k\tau_\beta(g_\alpha)\right)
  \right.\nonumber\\
&&\mbox{\hspace*{1cm}}\left.+(-1)^{n-k}\bbbone_p\otimes {\rm ad}_{\chi
  A_p\chi}^{(n-k)}(G_\alpha)\otimes\varphi\left((-i\partial_u)^ke^{\beta
  u/2}\tau_\beta(g_\alpha)\right)\right\}.\ \ \ \ \ \ \ \ \ \ 
\label{58}
\end{eqnarray}
Note that 
\begin{equation}
{\rm ad}_{\chi A_p\chi}^{(1)}(H_p)= \chi(H_p+W)\chi,
\label{56'}
\end{equation}
with 
\begin{equation}
W=-\frac{1}{2}\left(x\cdot\nabla
  v+2v\right)=\frac{1}{2}\left(\frac{\rho'(|x|)}{|x|^\mu}+(1-\mu)\frac{\rho(|x|)}{|x|^{1+\mu}}\right),
\label{57}
\end{equation}
and the choice of $\mu$, $\rho$ given in \fer{2} implies that
\begin{equation}
W\geq 0.
\label{59}
\end{equation}
In Appendix \ref{proofcpropsection} we prove the following Proposition.

\begin{proposition}
\label{Cprop}
$(C_n,\Lambda,\dom)$ are GJN triples, for $n=1,2,3$, and the estimates \fer{41}
  -\fer{40'} are satisfied. 
\end{proposition}
This shows that with the choice of operators introduced in this section,
Theorems \ref{virialthm} and \ref{virialthm'} apply.

\setcounter{equation}{0}
\section{Proof of Theorem \ref{nobifthm'}}
\label{proof1section}
1) \ Set $\widetilde I_1=i[I,A_f]$, where $A_f$ and $I$ are given in \fer{af},
   \fer{liouvillian}. For
   $\psi\in\dom(N^{1/2})$, we have 
\begin{equation*}
\left|\scalprod{\psi}{\widetilde I_1\psi}\right|\leq
\left|\scalprod{\psi}{\widetilde I_1\Pbar_\Omega\psi}\right| +
\left|\scalprod{\psi}{\Pbar_\Omega\widetilde I_1\pomega\psi}\right|
\leq 2\|\widetilde I_1 N^{-1/2}\Pbar_\Omega\|\|\psi\|\ \|N^{1/2}\psi\|.
\end{equation*}
This shows that in the sense of quadratic forms on $\dom(N)$,  
$
\lambda \widetilde I_1 \geq -cN-\frac{\lambda^2 k}{c},
$
for any $c>0$, where $k=\|\widetilde I_1N^{-1/2}\Pbar_\Omega\|^2\leq
k'\sum_\alpha\|\partial_u\tau_\beta (g_\alpha)\|^2_{L^2}\leq
k'(1+1/\beta)$. Let $\widetilde C_1=i[L_\lambda,A_f]=N+\lambda\widetilde I_1$
and choose $c=1/2$. Then we find 
$\widetilde C_1\geq \frac{1}{2}N -k\lambda^2$, in the sense of forms on
$\dom(N)\subset\dom (\widetilde C_1)$,
and from Theorem \ref{virialthm} 
\begin{equation*}
0=\lim_{\alpha\rightarrow 0}\scalprod{\psi_{\alpha}}{\widetilde C_1\psi_{\alpha}}\geq
\frac{1}{2}\lim_{\alpha\rightarrow 0}\|N^{1/2}\psi_\alpha\|^2
-k\lambda^2\|\psi\|^2,
\end{equation*}
where $\psi$ is an eigenvector of $L_\lambda$, and $\psi_\alpha$ its
regularization. It follows that 
\begin{equation*}
\lim_{\alpha\rightarrow 0}\|N^{1/2}\psi_\alpha\|^2\leq k\lambda^2\|\psi\|^2,
\end{equation*}
which tells us that $\psi\in\dom(N^{1/2})$, and that $\|N^{1/2}\psi\|\leq k
|\lambda|\, \|\psi\|$. \\
\indent
2)\  In what follows, the constants $k, k_1$, $\lambda_1$, $\lambda_2$  are
independent of $\beta\geq \beta_0$, where $\beta_0>0$ is arbitrary but
fixed. In the course of the proof we will impose several conditions on the
parameters $\epsilon, \lambda,\theta$ which are collected in
\fer{bcondition}. We adopt the notation of Subsection \ref{sub1}.\\ 
\indent
On $\dom(N)\subset \dom(C_1)$, we define the operator 
\begin{equation*}
B=C_1+i[L_\lambda,A_0].
\end{equation*}
Recall that $A_0$ is defined in
\fer{17}, that we write $\repsilonbar=\Pibar\repsilon$, and that 
$\pomega I\pomega=\pomega I_1\pomega=0$. Using that 
$P_\Omega=\Pi+\Pbar_0\otimes P_\Omega$, one finds that  
\begin{eqnarray}
P_\Omega BP_\Omega&=&\Pi B\Pi+\Pbar_0\otimes P_\Omega B \Pi+\Pi
B\Pbar_0\otimes P_\Omega +\Pbar_0\otimes P_\Omega B\Pbar_0\otimes
P_\Omega\nonumber\\
&=&2\theta\lambda^2\Pi I\repsilonbar^2I\Pi +\theta\lambda^2\left(
  \Pbar_0\otimes P_\Omega I\repsilonbar^2 I\Pi +\Pi I\repsilonbar^2 I
  \Pbar_0\otimes P_\Omega\right)\!\!,\ \ \ \ \ \ \label{28}\\
P_\Omega B \Pbar_\Omega&=&\lambda P_\Omega I_1 \Pbar_\Omega +\theta\lambda
\Pi I\repsilonbar^2 L_0 \Pbar_\Omega+\theta\lambda^2 \Pi I\repsilonbar^2
I\Pbar_\Omega,
\label{29}\\
\Pbar_\Omega B P_\Omega&=&\lambda \Pbar_\Omega I_1 P_\Omega +\theta\lambda
\Pbar_\Omega  L_0\repsilonbar^2I \Pi+\theta\lambda^2 \Pbar_\Omega I\repsilonbar^2
I \Pi,\label{29'}\\
\Pbar_\Omega B\Pbar_\Omega&=&\Pbar_\Omega N+\Pbar_\Omega\left(\lambda
I_1-\theta\lambda^2 (I\Pi I\repsilonbar^2+\repsilonbar^2 I\Pi I)\right)\Pbar_\Omega.\nonumber
\end{eqnarray}
From the estimates $\|\Pbar_\Omega N^{-1/2}I_1N^{-1/2}\Pbar_\Omega\|\leq k$, $\|I\Pi I\|\leq k$,
$\|\repsilonbar^2\|\leq \epsilon^{-2}$, we see that 
there is some constant $\lambda_1<\infty$ (independent of
$\lambda,\epsilon,\theta$), s.t.  
\begin{equation}
\Pbar_\Omega B\Pbar_\Omega \geq \frac{1}{2}\Pbar_\Omega,
\label{30}
\end{equation}
provided
\begin{equation}
 |\lambda|, \ \frac{\theta\lambda^2}{\epsilon^2}<\lambda_1,
\label{34}
\end{equation}
see also \fer{bcondition}. 
Using the estimates 
\begin{equation*}
\mbox{$\|I\repsilonbar^2I\Pi\|, \|\Pi I\repsilonbar^2 I\|\leq\epsilon^{-2} k$
  \ and \ $\|P_\Omega I_1\|,\| I_1P_\Omega\|\leq k$,}
\end{equation*}
where $k$ is independent of the parameters $\lambda,\theta,\epsilon$, we
arrive at the following lower bound. For any $\phi\in\dom(N)$ and some
$k_1<\infty$ 
\begin{eqnarray}
\av{B}_{\phi}&\geq& 2\theta\lambda^2\av{\Pi I\repsilonbar^2
  I\Pi}_\phi +\frac{1}{2}\|\Pbar_\Omega \phi\|^2\nonumber\\
&& -k_1\frac{\theta\lambda^2}{\epsilon^2}\|\Pi\phi\|\
  \|\Pbar_0\otimes\pomega \phi\|-k_1|\lambda| \ \|P_\Omega \phi\|\
  \|\Pbar_\Omega\phi\|\nonumber\\ 
&&-\left(2\theta|\lambda| +2k_1\frac{\theta \lambda^2}{\epsilon}\right)
  \|\repsilonbar I\Pi\phi\|\ \|\Pbar_\Omega\phi\|.
\label{lastline}
\end{eqnarray}
Clearly,  
$
\|\repsilonbar I\Pi\phi\|\ \|\Pbar_\Omega\phi\|\leq\delta\av{\Pi
  I\repsilonbar^2 I\Pi}_\phi+\delta^{-1} \|\Pbar_\Omega\phi\|^2$, for any
$\delta>0$. 
Choosing appropriate values of $\delta$, we bound the last line in
\fer{lastline} from below by
\begin{equation*}
-\theta\lambda^2 \av{\Pi I\repsilonbar^2 I\Pi}_\phi 
-4\left(\theta+k_1^2\frac{\theta\lambda^2}{\epsilon^2}\right)
\|\Pbar_\Omega\phi\|^2,
\end{equation*}
and it follows that
\begin{eqnarray}
\av{B}_{\phi}&\geq& \frac{\theta\lambda^2}{\epsilon}\gamma
\|\Pi\phi\|^2+\left(\frac{1}{2}  
-4\theta-4k_1^2\frac{\theta\lambda^2}{\epsilon^2}\right)\|\Pbar_\Omega
  \phi\|^2\nonumber\\ 
&& -k_1\frac{\theta\lambda^2}{\epsilon^2}\|\Pi\phi\|\
  \|\Pbar_0\otimes\pomega \phi\|-k_1|\lambda| \ \|P_\Omega \phi\|\
  \|\Pbar_\Omega\phi\|,
\label{lastline'}
\end{eqnarray}
where we have used \fer{fgrc1} and hence assumed that $0<\epsilon<\epsilon_0$.
Using that $\|P_\Omega\phi\|\leq \|\Pi\phi\|+\|\Pbar_0\otimes P_\Omega\phi\|$,
we estimate the two terms in the last line on the r.h.s. of \fer{lastline'} as 
\begin{eqnarray*}
-k_1|\lambda|\ \|P_\Omega\phi\|\ \|\Pbar_\Omega\phi\|
&\geq& -\frac{1}{4}\|\Pbar_\Omega\phi\|^2
-8\lambda^2 k_1^2\left( \|\Pi\phi\|^2+\|\Pbar_0\otimes
  P_\Omega\phi\|^2\right),\\
-k_1 \frac{\theta\lambda^2}{\epsilon^2} \|\Pi\phi\|\ \|\Pbar_0\otimes
P_\Omega\phi\| 
&\geq& -\frac{1}{2} \frac{\theta\lambda^2}{\epsilon}\gamma \|\Pi\phi\|^2 -
2k_1^2 \frac{\theta\lambda^2}{\epsilon^3}\gamma \|\Pbar_0\otimes
P_\Omega\phi\|^2. 
\end{eqnarray*}
Using these two estimates in \fer{lastline'}, we arrive at
\begin{equation}
\av{B}_\phi\geq \lambda^2\left(\frac{1}{2}\frac{\theta}{\epsilon}
  \gamma-8k_1^2\right)  
\|\Pi\phi\|^2 
-2\lambda^2k_1^2 \left(4+ \frac{\theta}{\epsilon^3\gamma}\right)
\|\Pbar_0\otimes P_\Omega\phi\|^2,
\label{fix1}
\end{equation}
where we require the condition
\begin{equation}
\frac{1}{4} -4\theta-4k_1^2\frac{\theta\lambda^2}{\epsilon^2}\geq 0,
\label{acondition}
\end{equation}
which guarantees that the contribution of the term in \fer{lastline'} which is
proportional 
to $\|\Pbar_\Omega\phi\|^2$ is non-negative, and can hence be
dropped. \fer{acondition} is satisfied if \fer{bcondition} holds.  
Let $\phi=\psi_{\alpha}\in \dom(N)$  be the regularization of the eigenvector
$\psi$ as defined in Theorem \ref{virialthm}. Then it follows from \fer{fix1}
that 
\begin{equation}
0=\lim_{\alpha\rightarrow
  0}\av{B}_{\psi_{\alpha}}\geq
\kappa_1\|\Pi\phi\|^2 -\kappa_2\|\Pbar_0\otimes P_\Omega\phi\|^2,
\label{32}
\end{equation}
where 
\begin{eqnarray}
\kappa_1&=&\frac{1}{2}\frac{\theta}{\epsilon}\gamma -8k_1^2 > k_1^2,
\label{kappa_1}\\
\kappa_2&=&2 k_1^2\left(4+\frac{\theta}{\gamma\epsilon^3}\right)>0.
\label{kappa_2}
\end{eqnarray}
The lower bound \fer{kappa_1} is a consequence of
$\epsilon<\frac{\theta\gamma}{18k_1^2}$, see \fer{bcondition}. \\
\indent
From \fer{regbnd}, we find that 
\begin{equation*}
\|\Pi\psi\|\geq\|\psi\|-\|\Pbar_\Omega\psi\|-\|\Pbar_0\otimes
P_\Omega\psi\| \geq (1-k|\lambda|)\|\psi\| -\|\Pbar_0\otimes
P_\Omega\psi\|.
\end{equation*}
Thus there is a positive constant $\lambda_2$ (independent of $\epsilon,
\lambda,\theta$) s.t. if $0<|\lambda|<\lambda_2$ then
$\|\Pi\psi\|\geq\frac{1}{2} \|\psi\| -\|\Pbar_0\otimes P_\Omega\psi\|$. Thus we
get from \fer{32} 
\begin{equation}
\| \Pbar_0\otimes P_\Omega\psi\|\geq
\frac{1}{2}\frac{1}{1+(\kappa_2/\kappa_1)^{1/2}}\|\psi\|.
\label{32.1}
\end{equation}
Consequently, under the conditions that 
\begin{equation}
0<|\lambda|<\min\left\{\lambda_1, \lambda_2, 
 \frac{\epsilon}{\sqrt{\theta}}\left(\sqrt{\lambda_1}+
 \frac{1}{4\sqrt{2}k_1}\right)\right\},\ \theta<\frac{1}{32},\
 \epsilon<\min\left\{\frac{\theta\gamma}{18k^2_1}, \epsilon_0\right\},
\label{bcondition}
\end{equation}
we obtain 
\begin{equation}
\|\Pbar_0\otimes P_\Omega\psi\|\geq \frac{1}{2}\frac{1}{1+\sqrt{2(4+\frac{\theta}{\gamma\epsilon^3})}}
\label{thebound}
\end{equation}
Choose for instance $\theta=1/100$, $\epsilon=\min\{\frac{\gamma}{2000k_1^2}, \epsilon_0\}$. Then
\fer{bcondition} holds provided $0<|\lambda|<k\gamma$, for some $k$
independent of $\beta$ provided that $\beta\geq\beta_0$, with $\beta_0>0$
arbitrary but fixed. For large $\beta$, the r.h.s. of \fer{thebound} behaves
like $\gamma^2$.
\hfill $\blacksquare$

\setcounter{equation}{0}
\section{Proof of Theorem \ref{regintthm}}
\label{proof2section}

The $\sigma_{t,\lambda}$-invariant normal states on $\m_\beta$ are in
one-to-one 
correspondence with the normalized vectors in the span of ${\cal P}\cap \ker
L_\lambda$ (see Section \ref{achsection}). Our task is
to show that ${\cal P}\cap \ker L_\lambda$ equals the set \fer{res1}. \\
\indent
In this section, we will always deal with the cutoff interaction
(determined by $G_{\alpha,J}$)  but we shall drop the subscript $_J$ in the
notation.

\subsection{Reduction of the Liouvillian}
We define the projection $p=p_{J_d}+p_{J_c}$, where $p_{J_d}, p_{J_c}$ are the
projections corresponding to the discrete and continuous modes in $J_d$ and
$J_c$, respectively; (see also \fer{n1}). Setting $\pbar=\bbbone_p-p$,
$P=p\otimes p\otimes\bbbone_f$, we decompose $\Pbar=\bbbone-P$ as
$
\Pbar=P^l+P^r+P^0
$, where 
\begin{equation}
P^l=p\otimes\pbar\otimes\bbbone_f\ \ \ P^r=\pbar\otimes p\otimes\bbbone_f,\ \
\ P^0=\pbar\otimes\pbar\otimes\bbbone_f.
\label{n2}
\end{equation}
It is easy to verify that the (regularized) Liouvillian $L_\lambda$, defined in
\fer{liouvillian}, is reduced by the decomposition
\begin{equation*}
\h=\ran P\oplus \ran P^l\oplus\ran P^r\oplus \ran P^0,
\end{equation*}
and that 
\begin{equation}
L_\lambda\upharpoonright \ran P^0=L_0\upharpoonright\ran P^0.
\label{sharp'}
\end{equation}
From the definition, \fer{J}, of the modular conjugation $J$, it follows that 
\begin{equation}
JP^l=P^rJ.
\label{intertwine}
\end{equation}
Because every $\psi\in{\cal P}$ satisfies $J\psi=\psi$, \fer{intertwine}
implies that ${\cal P}\cap\ran P^l={\cal P}\cap\ran P^r=\{0\}$, and,
consequently, we have that 
\begin{equation}
{\cal P}\cap\ker L_\lambda ={\cal P}\cap\left( \ker L_\lambda\upharpoonright
  \ran P^0 \cup \ker L_\lambda\upharpoonright\ran P\right).
\label{n2'}
\end{equation}
We prove in the next section that $\ker L_\lambda\upharpoonright
\ran P =\{0\}$, which, together with \fer{n2'} and \fer{sharp'}, shows that
${\cal P}\cap \ker L_\lambda$ is given by the subspace defined in \fer{res1}.

\subsection{The kernel of $L_\lambda\upharpoonright\ran P$} 

\begin{theorem}
\label{keronP}
Given any $0<\beta<\infty$ and any $r>0$ there is a $\lambda_0(\beta,r)>0$
s.t. if 
$0<|\lambda|<\lambda_0(\beta,r)$ then $\ker L_\lambda \upharpoonright\ran
P=\{0\}$. Here, $\lambda_0(\beta,r)\geq k\gamma^2r$, for some $k$ independent
of $\beta,r$. The constant $\gamma$ is given in \fer{gamma'}. 
\end{theorem}

{\it Proof.\ } We use the notation of Subsection \ref{sub2} and write $L$ for
$L_\lambda$. In the spirit of the positive commutator method outlined in
Section \ref{pcmethodsection}, we want to
establish a lower bound on the expectation value
$\scalprod{\psi}{(C_1+i[L,A_0])\psi}$ (see also \fer{56}), where $\psi$ is a
(hypothetical) eigenvector of $L$ in $\ran P$. \\
\indent
Using the relative bound 
\begin{equation}
 \pm\lambda I_1\geq -cN-\frac{\lambda^2}{c},\ \ \ \forall c>0,
\label{60}
\end{equation}
with $c=1/10$, and \fer{59}, we obtain a lower bound (always
in the sense of quadratic forms on $\dom$)
\begin{eqnarray}
\lefteqn{
C_1+i[L,A_0]}\nonumber\\
&\geq&\chi^2
H_p\otimes\bbbone_p+\bbbone_p\otimes\chi^2H_p+\frac{9}{10}N+i[L,A_0]-\frac{\lambda^2}{10}\nonumber\\
&\geq&P\left(\chi^2H_p\otimes\bbbone_p\otimes P_\Omega
  +\bbbone_p\otimes\chi^2H_p\otimes
  P_\Omega+\frac{9}{10}\Pbar_\Omega+i[L,A_0]-
\frac{\lambda^2}{10}\right)P\nonumber\\  
&&+\Pbar i[L,A_0]P+Pi[L,A_0]\Pbar+\Pbar i[L,A_0]\Pbar
-\frac{\lambda^2}{10}\Pbar\nonumber\\
&=:& PM_0P+M_1,
\label{61}
\end{eqnarray}
where the bounded operators  $M_0$ and $M_1$ are given by
\begin{eqnarray}
M_0&:=&p_{J_c}H_p\otimes\bbbone_p\otimes P_\Omega+\bbbone_p\otimes
  p_{J_c}H_p\otimes P_\Omega +\frac{9}{10}\Pbar_\Omega \nonumber\\
&&\ \ \ +i[L,A_0] -\frac{\lambda^2}{10}, \label{62}\\
M_1&:=& \Pbar i[L,A_0] P +Pi[L,A_0] \Pbar +\Pbar i[L,A_0]\Pbar
  -\frac{\lambda^2}{10} \Pbar.\label{62'}
\end{eqnarray}
\indent
The difficult part of the proof of Theorem \ref{regintthm} is contained in the
following two propositions. 
\begin{proposition}
\label{propspecloc}
Suppose Proposition \ref{fgrprop} holds and that the parameters satisfy
\begin{equation}
0<|\lambda|<\min\left(1,\sqrt{r},\frac{\epsilon}{3\sqrt{\theta} k},
 \frac{\epsilon}{\sqrt k}\right), \ 0<\epsilon<\min
 (5\theta\gamma,\epsilon_0), \ 0<\theta<r/32, 
\label{n2}
\end{equation}
where $k$ is a constant depending on the interaction, but not on any of the
parameters $\epsilon, \lambda, \theta$, nor on $\beta$ (for $\beta\geq
\beta_0$, with $\beta_0>0$ fixed). Then there is an interval $\Delta$
around zero such that
\begin{equation}
PE^0_\Delta M_0 E_\Delta^0P\geq \frac{\theta\lambda^2}{\epsilon}\gamma
E_\Delta^0P,
\label{n3}
\end{equation}
where $E_\Delta^0=E_\Delta(L_0)$ is the spectral projection of $L_0$ onto
$\Delta$. 
\end{proposition}
\begin{proposition}
\label{propotherloc}
Assume that the conditions of Proposition \ref{propspecloc} are satisfied and
that 
\begin{equation}
|\lambda|<\frac{\gamma}{k}\min\left(1,\epsilon,
 \theta/\epsilon\right).
\label{n4}
\end{equation}
If $\psi\in\ran P$ is an eigenvector of $L_\lambda$ then 
\begin{equation}
\scalprod{\psi}{M_0\psi}\geq \frac{1}{2}\frac{\theta\lambda^2}{\epsilon}\gamma
\|\psi\|^2.
\label{n5}
\end{equation}
\end{proposition}
\ \indent
We may choose the parameters as $\lambda=\widetilde \lambda\gamma^2$,
$\epsilon=\widetilde\epsilon\gamma$, with $\widetilde\lambda$,
$\widetilde\epsilon$ and $\theta$ 
independent of the inverse temperature $\beta$. Conditions \fer{n2}
and \fer{n4} are satisfied provided $0<|\widetilde\lambda|<kr$, for some $k$
independent of $\beta$ and $r$. \\
\indent
Theorem \ref{keronP} is now proven as follows. Assume that $\psi\in\ran P$ is
an 
eigenvector of $L_\lambda$, and let $\psi_\alpha$ be the
family of approximate eigenvectors  given in Theorem
\ref{virialthm'}. From \fer{61}, \fer{n5}, and using that  
$\scalprod{\psi}{M_1\psi}=0$, we obtain 
\begin{equation*}
0=\lim_\alpha\scalprod{\psi_\alpha}{(C_1+i[L,A_0])\psi_\alpha}\geq
\frac{1}{2}\frac{\theta\lambda^2}{\epsilon}\gamma\|\psi\|^2,
\end{equation*}
which is a contradiction, because the r.h.s. is strictly positive. \\

{\it Proof of Proposition \ref{propspecloc}.\ } Pick $\Delta$ such that 
\begin{equation}
|\Delta|<\frac{1}{2} \min\Big\{ |E(m)-E(n)|\ \big|\  m,n\in
J_d, E(m)\neq E(n)\Big\}.
\label{deltainterval}
\end{equation}
The Hilbert space $\ran E_\Delta^0P$ has the decomposition
\begin{equation}
\ran E_\Delta^0P=\ran P\Pi\oplus \ran E_\Delta^0 P\Pibar,
\label{n6}
\end{equation}
where, we recall, $\Pi= P_0\otimes P_\Omega$ is the projection onto the
kernel of $L_0$. We analyze the spectrum of the operator $PE_\Delta^0 M_0E_\Delta^0P$ on $\ran E_\Delta^0P$
using the Feshbach method. For details on the Feshbach method, we refer
to [BFS]. Let $m$ be a number in 
the resolvent set of $\Pibar PE_\Delta^0M_0E_\Delta^0P\Pibar$ (viewed as an
operator on $\ran E_\Delta^0P\Pibar$). The Feshbach map $F_{\Pi,m}$ applied to
the operator $PE_\Delta^0M_0E_\Delta^0P$ is defined as
\begin{eqnarray}
\lefteqn{
F_{\Pi,m}(PE_\Delta^0M_0E_\Delta^0P)}\label{70}\\
&=&\Pi\left( PM_0P-PM_0\Pibar P E_\Delta^0 \left(\Pibar PE_\Delta^0
    M_0E_\Delta^0 P\Pibar-m\right)^{-1} E_\Delta^0 P\Pibar M_0P\right)\Pi,
\nonumber
\end{eqnarray}
and has the following property of
{\it isospectrality}: Let $\sigma$ and $\rho$ denote the spectrum
and the resolvent set of an operator. Then 
\begin{eqnarray}
\lefteqn{
z\in \sigma(PE_\Delta^0 M_0E_\Delta^0P)\cap\rho(\Pibar
PE_\Delta^0M_0E_\Delta^0P\Pibar)}\label{71}\\
&&\Longleftrightarrow z\in\sigma(F_{\Pi,m}(PE_\Delta^0M_0E_\Delta^0P))\cap\rho(\Pibar PE_\Delta^0M_0E_\Delta^0P\Pibar).
\nonumber
\end{eqnarray}
The point of the Feshbach method is that it can be easier to analyze the
spectrum of the operator \fer{70} than 
the one of $PE_\Delta^0M_0E_\Delta^0P$, because the operator \fer{70} acts on
the smaller space $\ran P\Pi$. \\
\indent
Let us examine the diagonal blocks of $PE_\Delta^0 M_0E_\Delta^0P$ in the
decompositon \fer{n6}. It is readily verified that 
\begin{equation}
\Pi PM_0 P \Pi =2\theta\lambda^2 \Pi P I\repsilonbar^2 IP\Pi 
-\frac{\lambda^2}{10}P\Pi.
\label{n7}
\end{equation}
Because of \fer{deltainterval}, 
$E_\Delta^0\Pbar_0 \ p_{J_d}\otimes p_{J_d}\otimes P_\Omega=0$. Hence 
\begin{equation*}
E_\Delta^0 P\Pibar=E_\Delta^0 P \Pbar_\Omega + E_\Delta^0 P\left(
  p_{J_d}\otimes p_{J_c} +p_{J_c}\otimes p_{J_d} + p_{J_c}\otimes
  p_{J_c}\right)\otimes P_\Omega.
\end{equation*}
Set $Q_1=E_\Delta^0 P \Pbar_\Omega$ and let $Q_2$ be the other projection on
the right side. $PE_\Delta^0 M_0E_\Delta^0P$ is diagonal in this
decomposition of $\ran E_\Delta^0 P\Pibar$. We have the estimates 
\begin{equation*}
Q_1 M_0 Q_1 \geq
\left(\frac{9}{10}-\frac{\lambda^2}{10}-
  k\frac{\theta\lambda^2}{\epsilon^2}\right) Q_1, \ \ 
Q_2 M_0 Q_2 \geq \left(r-\frac{\lambda^2}{10}\right) Q_2,
\end{equation*}
where $k=2\|I\Pi I\|$ and we used that $p_{J_c}H_p\geq rp_{J_c}$. Consequently, we obtain the lower bound
\begin{equation}
P\Pibar E_\Delta^0  M_0 E_\Delta^0 \Pibar P\geq \min\left( 
\frac{9}{10}-\frac{\lambda^2}{10} -k\frac{\theta\lambda^2}{\epsilon^2},
r-\frac{\lambda^2}{10}\right) E_\Delta^0 \Pibar P\geq \frac{r}{2} E_\Delta^0
\Pibar P,
\label{n10}
\end{equation}
due to \fer{n2}. It follows that any $m<r/4$ is in the resolvent set of the
operator 
$P\Pibar E_\Delta^0 M_0 E_\Delta^0\Pibar P$ and 
\begin{equation}
\left\| \left( P\Pibar E_\Delta^0 M_0 E_\Delta^0\Pibar P-m\right)^{-1}\right\|
\leq 4/r.
\label{n11}
\end{equation}
We show now that
\begin{equation}
F_{\Pi,m}(PE_\Delta^0M_0E_\Delta^0P)\geq
\frac{\theta\lambda^2}{\epsilon}\gamma \ E_\Delta^0 P,
\label{n12}
\end{equation}
uniformly in $m<r/4$, provided \fer{n2} is satisfied, and where the Feshbach
map was introduced in \fer{70}. The bound \fer{n12} and the
isospectrality property \fer{71} of the Feshbach map imply \fer{n3}. \\
\indent
We complete the proof of the proposition by showing \fer{n12}. From \fer{n11}
it follows that for any $\psi$ and $m<r/4$ 
\begin{eqnarray}
\lefteqn{
\scalprod{\psi}{
\Pi PM_0\Pibar P E_\Delta^0 \left(\Pibar PE_\Delta^0
    M_0E_\Delta^0 P\Pibar-m\right)^{-1} E_\Delta^0 P\Pibar
  M_0P\Pi\psi}}\nonumber\\
&&\leq \frac{4}{r}\|\Pibar P E_\Delta^0 M_0 P \Pi \psi\|^2\nonumber\\
&&\leq \frac{8\theta^2\lambda^2}{r}(1+k\lambda^2/\epsilon^2) \scalprod{\psi}{
  P\Pi I\repsilonbar^2 I \Pi P\psi},
\label{n13}
\end{eqnarray}
where we estimate $\Pibar P E_\Delta^0 M_0 \Pi P=\theta\lambda \Pibar P
E_\Delta^0 L\repsilonbar^2 I\Pi P$ as
\begin{equation*}
\|\Pibar P E_\Delta^0 M_0 \Pi P\psi\|\leq \theta
|\lambda|(1+k|\lambda|/\epsilon)\| \repsilonbar I\Pi P\psi\|,
\end{equation*}
with $k=\|I P(N\leq 2)\|$. Taking into account \fer{n7} and Proposition
\ref{fgrprop}, we obtain the estimate
\begin{equation}
F_{\Pi,m}(PE_\Delta^0M_0E_\Delta^0P)\geq
  2\frac{\theta\lambda^2}{\epsilon}\gamma\ \left(
  1-\frac{4\theta}{r}(1+k\lambda^2/\epsilon^2)\right) \Pi P
  -\frac{\lambda^2}{10}\Pi P.
\label{n14}
\end{equation}
The bound \fer{n12} follows from \fer{n14} and conditions \fer{n2}. This
finishes the proof of Proposition \ref{propspecloc}.\\

{\it Proof of Proposition \ref{propotherloc}.\ } Let $0\leq g\leq 1$ be a
smooth function with support in the interval $\Delta$, s.t. $g=1$ on the
interval $(-|\Delta|/4,|\Delta|/4)$, and
set $g_0=g(L_0)$. Since the interaction $I$ is relatively $N^{1/2}$-bounded, it
follows in a standard way (by using 
e.g. the functional calculus presented in Appendix \ref{appendix}) that  
\begin{eqnarray}
\|(1-g_0)\psi\|&\leq& k|\lambda|\ \|(N+1)^{1/2}\psi\|\leq k|\lambda|\
\|\psi\|, \label{n15}\\
\|(1-g_0)\Pbar_\Omega\psi\|&\leq& k|\lambda|\ \|N^{1/2}\Pbar_\Omega\psi\|\leq
k|\lambda|^2\ \|\psi\|, \label{n15'}
\end{eqnarray}
where we used \fer{regbnd} in the last line. Notice also that $\Pi(1-g_0)=0$
and that on $\ran (1-g_0)$ we have $|L_0|\geq |\Delta|/4$ hence 
$\|(1-g_0)\repsilon\|\leq 4/|\Delta|$. These estimates are used below without
explicit mention.  
We decompose
\begin{eqnarray}
PM_0 P&=& g_0 P M_0 Pg_0\nonumber\\
&&+2\RE (1-g_0) P M_0 P g_0\label{n16}\\
&&+(1-g_0) P M_0 P(1-g_0).\label{n17}
\end{eqnarray}
Proposition \ref{propspecloc} yields the bound
\begin{equation}
\scalprod{\psi}{g_0 P M_0 Pg_0\psi} \geq
\frac{\theta\lambda^2}{\epsilon}\gamma\ \|g_0\psi\|^2= 
\frac{\theta\lambda^2}{\epsilon}\gamma\ (1-k|\lambda|)^2\|\psi\|^2.
\label{n18}
\end{equation}
We estimate
\begin{equation}
2\RE (1-g_0) P M_0 Pg_0\geq 2\RE (1-g_0)P i[L,A_0] P g_0
-\lambda^2(1-g_0)g_0 P,
\label{n19}
\end{equation}
and since 
\begin{equation*}
(1-g_0)i[L,A_0]g_0=-\theta\lambda (1-g_0)\left( \lambda I\Pi I\repsilonbar^2
  -L\repsilonbar^2 I\Pi -\lambda\repsilonbar^2 I\Pi I\right)g_0
\end{equation*}
we conclude that 
\begin{equation}
2\RE \scalprod{\psi}{(1-g_0)PM_0Pg_0\psi} \geq
-k\frac{\theta\lambda^2}{\epsilon}\left(\frac{|\lambda|}{\epsilon}
+\frac{|\lambda|\epsilon}{\theta}\right) \|\psi\|^2.
\label{n20}
\end{equation}
Next, we have that 
\begin{equation*}
(1-g_0)M_0(1-g_0)\geq -\theta\lambda^2(1-g_0) \left(I\Pi I\repsilonbar^2
  +\repsilonbar^2 I\Pi I\right) (1-g_0) -\frac{\lambda^2}{10}(1-g_0)^2,
\end{equation*}
from which it follows that 
\begin{equation}
\scalprod{\psi}{(1-g_0) PM_0P(1-g_0)\psi} \geq
-k\frac{\theta\lambda^2}{\epsilon}
\left(\lambda^4\epsilon+\frac{\lambda^2\epsilon}{\theta}\right)\|\psi\|^2.
\label{n22}
\end{equation}
Collecting the bounds \fer{n18}, \fer{n20} and \fer{n22} we obtain 
\begin{equation*}
\scalprod{\psi}{PM_0P\psi}\geq \frac{\theta\lambda^2}{\epsilon}\gamma\
\left(1-k|\lambda|- \frac{k}{\gamma}\left( \frac{|\lambda|}{\epsilon}
    +\frac{|\lambda| \epsilon}{\theta}\right)\right) \|\psi\|^2,
\end{equation*}
and \fer{n5} follows from the conditions \fer{n4}. This completes the proof of
Proposition \ref{propotherloc} and of Theorem \ref{keronP}. \hfill
$\blacksquare$

\appendix

\setcounter{equation}{0}
\section{Appendix}
\label{appendix}

\subsection{Invariance of domains, commutator expansion}
\label{invdomsect}

The following two theorems are useful in our analysis.
\begin{theorem}{\bf (invariance of domain, [F]) }
\label{IDthm}
 Suppose $(X,Y,\dom)$ satisfies
    the GJN Condition, \fer{nc1}, \fer{nc2}. Then the unitary group generated by the selfadjoint
    $X$, $e^{itX}$, leaves $\dom(Y)$ invariant, and we have the estimate
\begin{equation}
\|Ye^{itX}\psi\|\leq e^{k|t|}\|Y\psi\|,
\label{80}
\end{equation}
for some $k\geq 0$, and all $\psi\in\dom(Y)$.
\end{theorem}

\begin{theorem}{\bf (commutator expansion, [F]) }
\label{CEthm}
 Suppose $\dom$ is a core
  for the selfadjoint $Y\geq\bbbone$. Let $X,Z$ be two symmetric operators on
  $\dom$, and define the symmetric operators ${\rm ad}_X^{(n)}(Z)$ on $\dom$
  by 
\begin{eqnarray*}
{\rm ad}_X^{(0)}(Z)&=&Z,\\
\scalprod{\psi}{{\rm ad}_X^{(n)}(Z)\psi}&=&i\left\{\scalprod{{\rm
      ad}_X^{(n-1)}(Z)\psi}{X\psi} -\scalprod{X\psi}{{\rm
      ad}_X^{(n-1)}(Z)\psi}\right\},
\end{eqnarray*}
for all $\psi\in\dom$, $n=1,\ldots,M$. We suppose that the triples
$({\rm ad}_X^{(n)}(Z),Y,\dom)$, $n=0,1,\ldots,M$, satisfy the
GJN Condition \fer{nc1}, \fer{nc2}, and that $X$ is selfadjoint,
$\dom\subset\dom(X)$, $e^{itX}$ leaves $\dom(Y)$ invariant, and that\fer{80}
holds. The we have on $\dom(Y)$
\begin{eqnarray}
e^{itX}Ze^{-itX}&=& Z-\sum_{n=1}^{M-1}\frac{t^n}{n!}
{\rm ad}_X^{(n)}(Z)\nonumber\\
&&-\int_0^tdt_1\cdots\int_0^{t_{M-1}} dt_M e^{it_MX}{\rm
  ad}_X^{(M)}(Z)e^{-it_MX}. 
\label{cmexp}
\end{eqnarray}
\end{theorem}

The following Lemma is a consequence of the above two theorems. 

\begin{lemma}
\label{invlemma}
Suppose $(X,Y,\dom)$ and $({\rm ad}_X^{(n)}(Y),Y,\dom)$ are GJN triples, for
  $n=1,\ldots,M$, some $M\geq 
  1$. Moreover, assume that in the sense of Kato on $\dom(Y)$:
  $\pm{\rm ad}_X^{(M)}(Y)\leq k X$, for some $k\geq 0$. For $\chi\in
  {\cal S}(\r)$, a smooth function of rapid decrease,  define
  $\chi(X)=\int\hat{\chi}(s)e^{isX}$, where $\hat{\chi}$ is the Fourier
  transform of $\chi$. Then $\chi(X)$ leaves $\dom(Y)$ invariant.
\end{lemma}

{\it Proof.\ }
For $R>0$, set $\chi_R(X)=\int_{-R}^R\hat{\chi}(s)e^{isX}$, then
$\chi_R(X)\rightarrow \chi(X)$ in operator norm, as $R\rightarrow
\infty$. From the invariance of domain theorem, we see that $\chi_R(X)$ leaves
$\dom(Y)$ invariant. Let $\psi\in\dom(Y)$, then using the commutator expansion
theorem above, we have
\begin{eqnarray}
\lefteqn{
Y\chi_R(X)\psi}\nonumber\\
&=&\chi_R(X)Y\psi +\int_{-R}^R\hat{\chi}(s)
    e^{isX}\left(e^{-isX} Y e^{isX}-Y \right)\psi\nonumber\\
&=&\chi_R(X)Y\psi-\int_{-R}^R\hat{\chi}(s)e^{isX}\left(\sum_{n=1}^{M-1}\frac{(-s)^n}{n!}{\rm
    ad}_X^{(n)}(Y)\right.\nonumber\\
&&\left.+(-1)^M\int_0^sds_1\cdots\int_0^{s_{M-1}}ds_Me^{-is_MX}{\rm
    ad}_X^{(M)}(Y) e^{is_MX}\right)\psi
\label{81}
\end{eqnarray}
The integrand of the $s$-integration in \fer{81} is bounded in norm by 
\begin{equation*}
k(|s|^M+1)\left(\|Y\psi\|+\|X\psi\|\right)\leq k(|s|^M+1)\|Y\psi\|,
\end{equation*}
where we used that $\|{\rm ad}_X^{(M)}(Y)e^{is_MX}\psi\|\leq
k \|Xe^{is_MX}\psi\|=k \|X\psi\|$. Since $\hat{\chi}$ is of rapid decrease, it
can be integrated against any power of $|s|$, and we conclude that the
r.h.s. of \fer{81} has a limit as $R\rightarrow\infty$. Since $Y$ is a closed
operator it follows that $\chi(X)\psi\in\dom(Y)$.\hfill $\blacksquare$\\

\begin{lemma}
\label{boundedcommlemma}
Let $\chi\in C_0^\infty(\r^3)$, $\chi=F^2\geq 0$. Suppose $(X,Y,\dom)$
satisfies the GJN condition and define $F(X), \chi(X)$ via the Fourier
transform as in Lemma \ref{invlemma}. Suppose $F(X)$ leaves $\dom(Y)$
invariant. Let $Z$ be a symmetric operator on $\dom$, s.t. for some $M\geq 1$,
and $n=0,1,\ldots,M$, the triples $({\rm ad}_X^{(n)}(Z),Y,\dom)$ satisfy the
GJN condition. Moreover, assume that the multi-commutators for $n=1,\ldots,M-1$
are bounded, and the $M$-th multicommutator is relatively $X$-bounded in the
sense of Kato on $\dom$: there is a $k<\infty$ s.t. $\forall \psi\in \dom$:
\begin{eqnarray*}
\|{\rm ad}_X^{(n)}(Z)\psi\|&\leq& k\|\psi\|,\ \ \ \ n=1,\ldots,M-1,\\
\|{\rm ad}_X^{(M)}(Z)\psi\|&\leq& k\|X\psi\|.
\end{eqnarray*}
Then the commutator $[\chi(X),Z]=\chi(X)Z-Z\chi(X)$ is well defined on $\dom$
and bounded: there is a $k<\infty$ s.t. $\|[\chi(X),Z]\psi\|\leq k\|\psi\|$,
$\forall \psi\in\dom$.
\end{lemma}
\ \\
\indent
{\it Proof.\ }
We write $F,\chi$ instead of $F(X), \chi(X)$. Since $F$ leaves $\dom(Y)$
invariant we have in the strong sense on $\dom(Y)$:
\begin{equation*}
[\chi,Z]=F[F,Z]+[F,Z]F.
\end{equation*}
We expand the commutator
\begin{eqnarray*}
[F,Z]&=&\int\hat{F}(s) e^{isX}\left( Z-e^{-isX}Ze^{isX}\right)\\
&=&\int\hat{F}(s)e^{isX}\left\{\sum_{n=1}^{M-1}\frac{s^n}{n!}{\rm
    ad}_X^{(n)}(Z)\right.\\
&&\ \ \ \left.+ \int_0^sds_1\cdots\int_0^{s_{M-1}}ds_M e^{-is_MX} {\rm
    ad}_X^{(M)}(Z) e^{is_MX}\right\}.
\end{eqnarray*}
Multiplying this from the left with $F$ (and noticing that $F$ commutes with
$e^{is_MX}$), we see immediately that $[F,Z]F$ is bounded. Next, for any
$\phi,\psi$ in the dense set $\dom$, we have the estimate
$
\left|\scalprod{\phi}{F[F,Z]\psi}\right|=\left|\scalprod{[F,Z]F\phi}{\psi}\right|\leq
k\|\phi\|\ \|\psi\|
$, 
hence
$\|F[F,Z]\psi\|=\sup_{0\neq\phi\in\dom}\|\phi\|^{-1}|\scalprod{\phi}{F[F,Z]\psi}|\leq
k\|\psi\|$.\hfill $\blacksquare$

\subsection{Proof of Proposition \ref{Aprop}}
\label{proofprop3.2section}

Before proving Proposition \ref{Aprop}, we show certain triples satisfy the
GJN conditions.

\begin{lemma}
\label{NClemma}
 The following triples satisfy the GJN Condition
  \fer{nc1}, \fer{nc2}:
\begin{eqnarray}
&&(H_p,\Lambda_p,C_0^\infty(\r^3)) \label{82}\\
&&(A_p,\Lambda_p,C_0^\infty(\r^3)) \label{83}\\
&&({\rm ad}_{H_p}^{(n)}(\Lambda_p), \Lambda_p,C_0^\infty(\r^3)),\ \
n=1,2. \label{84}
\end{eqnarray}
\end{lemma}

{\it Proof.\ }
For $\psi \in C_0^\infty(\r^3)$ we have 
$
\|H_p\psi\|^2\leq 2\|(-\Delta)\psi\|^2 +2\|v\|_\infty^2 \|\psi\|^2.
$
The first term on the r.h.s. is bounded from above by $2\|\Lambda_p\psi\|^2
 +4\RE\scalprod{\psi}{\Delta x^2\psi}$ and we have the estimate
 $\RE\scalprod{\psi}{\Delta x^2\psi} =\sum_{j=1}^3(\scalprod{\psi}{x_j\Delta
   x_j\psi} +\scalprod{\psi}{\partial_jx_j\psi})\leq
 \sum_{j=1}^3\|\partial_j\psi\|\, \|x_j\psi\|\leq \scalprod{\psi}{(-\Delta
   +x^2)\psi}$. 
Therefore, since $\Lambda_p\geq \bbbone$, it follows that 
that $\|H_p\psi\|^2\leq
k \|\Lambda_p\psi\|^2$. In a similar way it is simple to verify that $\pm{\rm
  ad}_{\Lambda_p}^{(1)}(H_p)\leq k\Lambda_p$. This shows that \fer{82}
satisfies the GJN conditions. The proof 
for \fer{83} is similarly easy.\\
\indent
{}From the above calculations, and ${\rm ad}_{H_p}^{(1)}(\Lambda_p)=-{\rm
  ad}_{\Lambda_p}^{(1)}(H_p)$, we see that $\|{\rm
  ad}_{H_p}^{(1)}(\Lambda_p)\psi\|\leq k\|\Lambda_p^{1/2}\psi\|$. Moreover, we
calculate 
\begin{eqnarray*}
\lefteqn{
{\rm ad}_{\Lambda_p}^{(1)}\left({\rm
    ad}_{H_p}^{(1)}(\Lambda_p)\right)}\\
&&=-\left(2x\cdot\nabla v
    +2\partial_mv_{mn}\partial_n +\partial_n v_{nmm} +4\Delta +4x^2\right)
    +{\rm h.c.},
\end{eqnarray*}
where $v_{klm}=\partial_k\partial_l\partial_m v$, etc. Since all the
derivatives of $v$ involved are bounded, the r.h.s. is again
relatively $\Lambda_p$-bounded, in the form sense on $C_0^\infty(\r^3)$. This
shows  
\fer{84} for $n=1$. For $n=2$, we calculate
\begin{equation}
{\rm ad}_{H_p}^{(2)}(\Lambda_p)=-\left(\nabla v\cdot\nabla v +2x\cdot\nabla v
  +2\partial_m v_{nm}\partial_n +\partial_m v_{mnn}+4\Delta\right) +{\rm h.c.}
\label{87}
\end{equation} 
Notice that 
${\rm ad}_{H_p}^{(2)}(\Lambda_p)$ is relatively $H_p$-bounded, hence
relatively $\Lambda_p$-bounded, in the sense of Kato on
$C_0^\infty(\r^3)$, because $\nabla v\cdot \nabla v$ and $x\cdot\nabla v$ are
bounded. Moreover, one can  
calculate ${\rm ad}_{\Lambda_p}^{(1)}\left({\rm
    ad}_{H_p}^{(2)}(\Lambda_p)\right)$, and see that it is $\Lambda_p$ form
bounded on $C_0^\infty(\r^3)$. This shows \fer{84} for $n=2$.\hfill
$\blacksquare$\\

{\it Proof of Proposition \ref{Aprop}.\ }
We start by showing that $\chi$ leaves $\dom(\Lambda_p)$ invariant. This
follows from Lemma \ref{invlemma} and the following two facts: firstly,
$(H_p,\Lambda_p, 
C_0^\infty(\r^3))$ and $({\rm ad}_{H_p}^{(n)},\Lambda_p,C_0^\infty(\r^3))$
satisfy the GJN Condition, for $n=1,2$ (see Lemma \ref{NClemma}), and
secondly,  $\pm{\rm
  ad}_{H_p}^{(2)}(\Lambda_p)\leq k H_p$, in the sense of Kato on
$\dom(\Lambda_p)$ (see \fer{87}). This shows that
$\chi:\dom(\Lambda_p)\rightarrow \dom(\Lambda_p)$, and consequently, $\chi
A_p\chi$ is well defined on $\dom(\Lambda_p)$. \\
\indent
We have in the strong sense on $\dom(\Lambda_p)$
\begin{equation}
\chi A_p \chi=\chi^2 A_p +\chi[A_p,\chi].
\label{88}
\end{equation}
Let us estimate each term on the r.h.s. separately. For $\psi\in
C_0^\infty(\r^3)$, 
\begin{eqnarray*}
\|\chi^2 A_p\psi\|&=&\|\chi^2(\nabla\cdot x
+3/2)\psi\|\leq\sum_n\|\chi^2\partial_n\|\ \|x_n\psi\|+\frac{3}{2}\|\psi\|\\
&\leq& k\sum_n\scalprod{\psi}{x^2_n\psi}^{1/2} +\frac{3}{2}\|\psi\|
\leq k \scalprod{\psi}{(-\Delta+x^2)\psi}^{1/2},
\end{eqnarray*}
where we used that $\|\chi^2\partial_n\|<\infty$. Consequently,  
\begin{equation}
\|\chi^2 A_p\psi\|\leq k\|\Lambda^{1/2}_p\psi\|.
\label{89}
\end{equation}
Next we have on $\dom(\Lambda_p)$
\begin{equation*}
\chi[A_p,\chi]=-\chi\int ds\ \hat{\chi}(s)\int_0^s ds_1 e^{is_1H_p}{\rm
  ad}_{A_p}^{(1)}(H_p)e^{-is_1H_p},
\end{equation*}
where ${\rm ad}_{A_p}^{(1)}(H_p)=H_p+W$, see \fer{56'}, and since
$\chi$ commutes with $e^{is_1H_p}$, we obtain the estimate
\begin{equation}
\|\chi[A_p,\chi]\|\leq \int ds\ |\hat{\chi}(s)| s \left(\|\chi
  H_p\|+\|W\|_\infty\right)<\infty.
\label{98'}
\end{equation}
It follows, together with with \fer{88} and \fer{89}, that
\begin{equation*}
\|\chi A_p\chi\psi\|\leq k\left( \|\Lambda_p^{1/2}\psi\| +\|\psi\|\right)\leq
2k\|\Lambda_p^{1/2}\psi\|.
\end{equation*}
This shows that the first GJN condition, \fer{nc1}, is satisfied for our
triple. Next, we write
\begin{equation}
\scalprod{\chi
  A_p\chi\psi}{\Lambda_p\psi}=\scalprod{A_p\chi\psi}{\Lambda_p\chi\psi}+R_1,
\label{90}
\end{equation}
where 
\begin{equation}
R_1=\scalprod{A_p\chi\psi}{[\chi,\Lambda_p]\psi}.
\label{91}
\end{equation}
Since $\chi\psi\in\dom(\Lambda_p)$, and $\Lambda_p$ is essentially
selfadjoint on $C_0^\infty(\r^3)$, there exists a sequence
$\{\varphi_n\}\subset C_0^\infty(\r^3)$ s.t. $\varphi_n\rightarrow\chi\psi$
and $\Lambda_p\varphi_n\rightarrow\Lambda_p\chi\psi$. Moreover
$\Lambda_p$ leaves $C_0^\infty(\r^3)$ invariant, so we have
\begin{eqnarray}
\scalprod{A_p\chi\psi}{\Lambda_p\chi\psi}&=&\lim_n\scalprod{A_p\chi\psi}{\Lambda_p\varphi_n}\nonumber\\
&=&\lim_n\left\{\scalprod{\Lambda_p\chi\psi}{A_p\varphi_n}+\scalprod{\chi\psi}{[A_p,\Lambda_p]\varphi_n}\right\},
\label{92}
\end{eqnarray}
and we calculate (strongly on $C_0^\infty(\r^3)$):
$i\,[A_p,\Lambda_p]=H_p+W-x\cdot\nabla v-2x^2$. Since $\chi(H_p+W-x\cdot\nabla
v)$ is bounded, and $\chi\psi\in\dom(x^2)$ (because it is in
$\dom(\Lambda_p)$), we get 
\begin{equation}
\fer{92}=\lim_n\scalprod{\Lambda_p\chi\psi}{A_p\varphi_n}+R_2,
\label{93}
\end{equation}
where we defined
\begin{equation}
R_2=\scalprod{\psi}{\chi(H_p+W-x\cdot\nabla v-2x^2)\chi\psi}.
\label{94}
\end{equation}
Next, it is not difficult to see that if $\psi\in C_0^\infty$ then
$\chi\psi\in\dom(\Lambda_p^2)$. Consequently,
we can move $A_p$ in \fer{93} to the left factor in the scalar product (recall
that $\dom(A_p)\supset\dom(\Lambda_p)$), perform the limit, and move $A_p$ back
  to the right factor. We then obtain
\begin{equation}
\scalprod{A_p\chi\psi}{\Lambda_p\chi\psi}=\scalprod{\Lambda_p\chi\psi}{A_p\chi\psi}+R_2=\scalprod{\Lambda_p\psi}{\chi
  A_p\chi\psi}+R_2-\overline{R}_1,
\label{95}
\end{equation}
where the bar denotes complex conjugate. Together with \fer{90} this gives
\begin{equation}
\scalprod{\chi A_p\chi\psi}{\Lambda_p\psi}-\scalprod{\Lambda_p\psi}{\chi
  A_p\chi\psi}= R_1+R_2-\overline{R}_1.
\label{96}
\end{equation}
Let us first consider $R_1$. We estimate
\begin{equation*}
|R_1|\leq \|A_p\chi\psi\| \ \|[\chi,\Lambda_p]\psi\|\leq
\big(\|A_p\psi\| +\|[A_p,\chi]\psi\|\big)\ \|[\chi,\Lambda_p]\psi\|.
\end{equation*}
It is clear that $\|A_p\psi\|\leq k\|\Lambda_p^{1/2}\psi\|$, and by the same
argument as the one leading to \fer{98'}, that $\|[A_p,\chi]\psi\|\leq
k\|\psi\|$ (use that $\chi=F^2\geq 0$, as in the proof of Lemma
\ref{boundedcommlemma}), so that 
\begin{equation}
|R_1|\leq k\|\Lambda_p^{1/2}\psi\|\ \|[\chi,\Lambda_p]\psi\|\leq
 k\|\Lambda_p^{1/2}\psi\| \big(\|\psi\|+\|[\chi,x^2]\psi\|\big),
\label{97}
\end{equation}
where we used in the second step that $[\chi,-\Delta]$ is bounded. Next, 
Lemma \ref{boundedcommlemma} (with $X=H_p, Z=x_n, Y=\Lambda_p, M=1$) shows
that $[\chi,x_n]$ is bounded, hence $\|[\chi,x^2]\psi\|\leq \sum_n
(k\|x_n\psi\|+ \|x_n[\chi,x_n]\psi\|)\leq k\|\Lambda_p^{1/2}\psi\|
+\sum_n\|x_n[\chi,x_n]\psi\|$. We write 
$x_n[\chi,x_n]\psi=[\chi,x_n]x_n\psi+ [x_n,[\chi,x_n]]\psi$. As above,
$\|[\chi,x_n]x_n\psi\|\leq k \|\Lambda_p^{1/2}\psi\|$, and using $\chi^2=F$,
we write
\begin{equation}
[x_n,[\chi,x_n]]= -2[F,x_n]^2+F[x_n[F,x_n]]+[x_n,[F,x_n]]F. 
\label{98}
\end{equation}
As above, $[F,x_n]$ is bounded, and a by now standard commutator expansion
shows that so are the other two terms in \fer{98}. We conclude that
$\|[\chi,x^2]\psi\|\leq k(\|\Lambda_p^{1/2}\psi\|+\|\psi\|)\leq
k\|\Lambda_p^{1/2}\psi\|$. Combining this with \fer{97} yields
\begin{equation}
|R_1|\leq k\|\Lambda_p^{1/2}\psi\|^2.
\label{99}
\end{equation}
Finally, let us obtain the same upper bound for $R_2$. Since
$\chi(H_p+W-x\cdot \nabla v)$ is bounded we only need to show that
$|\scalprod{\chi\psi}{x^2\chi\psi}|\leq k\|\Lambda_p^{1/2}\psi\|^2$. Using that
$[\chi,x_n]$ is bounded, we arrive at
\begin{equation*}
|\scalprod{\chi\psi}{x^2\chi\psi}|=\sum_n\|x_n\chi\psi\|^2\leq k\|\psi\|^2
 +k\sum_n\|x_n\psi\|^2\leq k\|\Lambda_p^{1/2}\psi\|^2,
\end{equation*}
which shows that 
\begin{equation}
|R_2|\leq k\|\Lambda_p^{1/2}\psi\|^2.
\label{100}
\end{equation}
Combining \fer{96}, \fer{99} and \fer{100} shows that $(\chi
A_p\chi,\Lambda_p,C_0^\infty(\r^3))$ satisfies the second GJN condition,
\fer{nc2}.\hfill $\blacksquare$\\

\subsection{Proof of Proposition \ref{Cprop}}
\label{proofcpropsection}
We give the following lemma without a proof, which is not difficult to find,
e.g. by using the results of Appendix \ref{invdomsect}.

\begin{lemma}
\label{solemma}
Let $v\in C^{p-1}(\r^d)$ be s.t. $x^\alpha \partial^\alpha v$ is bounded, for
 any 
 multi-index $|\alpha|\leq p-1$. Let $H=-\Delta+v$, which is  essentially
 selfadjoint on 
 $C_0^\infty(\r^d)$. Given a function $\mu\in C_0^\infty(\r)$, define
 $\mu(H)=\int\hat{\mu}(s)e^{isH}$. Then we have
\begin{equation}
\|\x^{\mp n}\mu(H)\x^{\pm n}\|\leq K(p,\mu), \ \ \ n=0,1,\ldots,p,
\label{101}
\end{equation}
where $K(p,\mu)$ is some finite constant.
\end{lemma}

\begin{lemma}
\label{reglemma}
The regularized $G_{\alpha,J}=(p_{J_d}+\mu) G_\alpha (p_{J_d}+\mu)$ satisfies the same bounds
\fer{dec} as $G_\alpha$. Moreover, ${\rm ad}_{\chi
  A_p\chi}^{(n)}(G_{\alpha,J})$ is bounded, $n=0,1,2,3$.
\end{lemma}

{\it Proof.\ } The first assertion follows easily from the fact that $\x^m
p_{J_d}\x^n$ is bounded, for all $m,n$, and from Lemma \ref{solemma} (use
$\x^n\mu=\x^n\mu\x^{-n}\x^n$). \\ 
\indent
In order to show boundedness of the multi-commutators, we treat a typical term
appearing in ${\rm ad}_{\chi A_p\chi}^{(3)}(G_{\alpha,J})$:
\begin{eqnarray*}
\lefteqn{
\chi A_p\chi G_{\alpha,J}\chi A_p\chi \chi A_p\chi}\\
&=&\chi A_p\x^{-1}\x\chi\x^{-1}\x G_{\alpha,J}\x^2 \x^{-2}\chi
A_p\chi\x\x^{-1}A_p\chi.
\end{eqnarray*}
Since $\chi A_p\x^{-1}$ and $\x^{-1} A_p\chi$ are bounded, we see from Lemmas
\ref{solemma} and the bound \fer{dec} (for $G_{\alpha,J}$), that the r.h.s. is bounded, provided
$\|\x^{-2}\chi A_p\chi\x\|<\infty$. To obtain the latter bound, it is enough
(due to due to Lemma \ref{solemma}) 
to show $\|\x^{-2}\chi A_p\x\|<\infty$, which in turn is proved by
writing 
\begin{equation*}
\x^{-2}\chi A_p\x=\frac{i}{2}\x^{-2}\chi\sum_n (x_n\x\partial_n+1/2),
\end{equation*}
and proceeding as in the proof of Lemma \fer{solemma}, by commuting $x_n\x$
through $\chi$ to the left. \hfill $\blacksquare$\\

{\it Proof of Proposition \ref{Cprop}}.\ \ The operator 
$\chi(H_p+W)\chi$ is bounded, hence relatively
$\Lambda_p$-bounded. We will use below the fact that $[\chi,A_p]$ is bounded,
which follows from Lemma
\ref{boundedcommlemma}, with $X=H_p$, $Z=A_p$, $Y=\Lambda_p$, $M=1$. 
 We have, in the strong sense on $\dom(\Lambda_p)$,
\begin{equation*}
{\rm ad}_{\chi A_p\chi}^{(2)}(H_p)=\chi[\chi^2H_p,A_p]\chi +[\chi W\chi,\chi
A_p\chi],
\end{equation*}
where the commutator in the first term is bounded, and the second term equals 
$\chi(W\chi^2A_p-A_p\chi^2 W)\chi$, which is easily seen to be
bounded, too (use Lemma \ref{solemma} together with the fact that $\chi
A_p\x^{-1}$ is bounded, and so is $W\x$). Next, we show that
\begin{equation}
{\rm ad}_{\chi A_p\chi}^{(3)}(H_p)=[\chi[\chi^2H_p,A_p]\chi,\chi A_p\chi]
+[[\chi W\chi,\chi A_p\chi],\chi A_p\chi]
\label{107}
\end{equation}
is bounded. The first term on the r.h.s. is the sum of two bounded operators
plus $\chi[\, [\chi^2H_p,A_p],\chi A_p\chi]\chi$, which can be written (by
commuting $\chi$ through $A_p$) as a bounded operator plus
\begin{equation}
\chi[\, [\chi^2H_p,A_p], A_p\chi^2]\chi.
\label{108}
\end{equation}
 Setting $\chi_1=\chi H_p$,
we expand $[\chi_1,A_p]=\chi_1' H_p+\int\hat{\chi}_1(s)\int_0^s
e^{i(s-s_1)H_p}We^{is_1H_p}$, and \fer{108} splits into two terms, the first
one, $\chi [\chi'_1H_p,A_p]\chi^4$, is bounded. To see boundedness of the
second term, write it as
\begin{equation*}
\int\hat{\chi}_1(s)\int_0^s \chi\left\{[e^{i(s-s_1)H_p}We^{is_1H_p},A_p]\chi^2
  +A_pe^{i(s-s_1)H_p}[W,\chi^2]e^{is_1H_p}\right\}\chi,
\end{equation*}
and use that $\x W$ is bounded. The second term in \fer{107} can be written as
a bounded operator plus $\chi [W\chi^2 A_p-A_p\chi^2 W,\chi^2 A_p]\chi$. The
latter term equals again a bounded operator plus $2\RE [W\chi^2A_p,\chi^2
A_p]$, which is easily seen to be bounded, by again noticing that $\x W$, and
 derivatives of $W$ multiplied by $\x^2$ are bounded.\\
\indent
We have thus shown that the multi-commutators appearing in \fer{56}
are bounded (hence $\Lambda_p$-bounded in the sense of Kato). Clearly, $N$ is
$\Lambda_f$ bounded, 
and Lemma \ref{reglemma}, together with the fact that
$\varphi((-i\partial_u)^k\tau_\beta(g_\alpha)))$ is relatively $N^{1/2}$-bounded,
shows that $I_n$ is $\Lambda_f$-bounded in the sense of Kato. Consequently,
condition  \fer{nc1}  is satisfied for $X=C_n$.  \\
\indent
Next, we verify that \fer{nc2} is satisfied. Let us start with the commutator
of $C_1$ with $\Lambda$. We need to show relative boundedness of $[\chi
(H+W)\chi,\Lambda_p]$ and $[I_1,\Lambda]$ (relative to $\Lambda_p$ and
$\Lambda_f$, respectively, in the sense of quadratic forms). Noticing that
$\Lambda_p=H_p-v+x^2$, we write $[\chi (H+W)\chi,\Lambda_p]$  as a sum of a
bounded  operator plus 
\begin{equation}
[\chi^2H_p,x^2]+[\chi W\chi,x^2].
\label{109}
\end{equation}
Now setting $\chi_1=\chi^2 H_p$, the first term in \fer{109} equals
$\sum_n(x_n[\chi_1,x_n]+[\chi_1,x_n]x_n)$, so for any $\psi\in C_0^\infty$,
\begin{equation}
|\scalprod{\psi}{[\chi_1,x^2]\psi}\leq k\sum_n \|x_n\psi\|\
 \|\psi\|\leq k\scalprod{\psi}{\Lambda_p\psi}.
\label{110}
\end{equation}
Next, 
\begin{eqnarray}
\lefteqn{|\scalprod{\psi}{[\chi W\chi,x^2]\psi}|}\label{111}\\
&&\leq 2|\scalprod{\psi}{\chi
  W[\chi,x^2]\psi}|\leq 2\sum_n|\scalprod{\psi}{(\chi Wx_n[\chi,x_n]+\chi
  W[\chi,x_n]x_n)\psi}|.
\nonumber
\end{eqnarray}
Commuting $x_n$ in the first term in the sum through $\chi$ to the left, one
  sees that  $|\scalprod{\psi}{\chi
  Wx_n[\chi,x_n]\psi}|\leq k(\|\psi\|^2+\|x_n\psi\|\ \|\psi\|)$, which is
  bounded from above by $k\scalprod{\psi}{\Lambda_p\psi}$ (proceed as in
  \fer{110}).  The second term in the sum in \fer{111} is estimated in the
  same way. 
This shows that \fer{109} is $\Lambda_p$-form-bounded.\\
\indent
Next, in order to show the relative bound on $[I_n,\Lambda]$, it is enough to
  show that $[{\rm ad}_{\chi A_p\chi}^{(n)}(G_\alpha),\Lambda_p]$ is
  relatively $\Lambda_p$-form-bounded, and that
\begin{equation*}
[\varphi((-i\partial_u)^n\tau_\beta(g_\alpha), \Lambda_f]
\end{equation*}
 is relatively
  $\Lambda_f$-form-bounded. The former bound is easily obtained from
  \fer{dec}, and the latter has been treated in subsection \ref{sub1}. This shows
  that $I_n$ are 
relatively $\Lambda$-form-bounded, hence also completing the proof that $C_1$
satisfies condition \fer{nc2}. \\
\indent
Next, we consider the commutator of $C_2$ with $\Lambda$. The only thing to
check is that $[{\rm ad}_{\chi A_p\chi}(H_p),\Lambda_p]$ is
$\Lambda_p$-form-bounded. This commutator can be written as a bounded operator
plus $[{\rm ad}_{\chi A_p\chi}^{(2)}(H_p),x^2]$, hence it suffices to show
that $x_n[{\rm ad}_{\chi A_p\chi}^{(2)}(H_p),x_n]$ is $\Lambda_p$-form-bounded
($n=1,2,3$). One shows that $[{\rm ad}_{\chi A_p\chi}^{(2)}(H_p),x_n]$ is
bounded, by simple estimates as above. Relative boundedness of $x_n[{\rm
  ad}_{\chi A_p\chi}^{(2)}(H_p),x_n]$ then follows easily (proceeding as in
\fer{110}). Consequently, \fer{nc2} is satisfied for $C_2$.\\
\indent
We now consider the commutator of $C_3$ with $\Lambda$, and it is enough to
show that $[{\rm ad}_{\chi A_p\chi}^{(3)}(H_p),x^2]$ is relatively
$\Lambda_p$-form-bounded. We write this commutator as $2\RE\,[{\rm ad}_{\chi
  A_p\chi}^{(2)}(H_p)\chi A_p\chi,x^2]$. Now we have  
\begin{equation*}
[{\rm ad}_{\chi A_p\chi}^{(2)}(H_p)\chi A_p\chi,x_n]=[{\rm ad}_{\chi
  A_p\chi}^{(2)}(H_p),x_n]\chi A_p\chi+ 
{\rm ad}_{\chi A_p\chi}^{(2)}(H_p)[\chi A_p\chi,x_n],
\end{equation*}
and it is clear that $[{\rm ad}_{\chi A_p\chi}^{(2)}(H_p)\chi
A_p\chi,x_n] \x^{-1}$ is bounded. Consequently, 
\begin{equation*}
\left|\scalprod{\psi}{[{\rm ad}_{\chi A_p\chi}^{(3)}(H_p),x^2]\psi}\right|\leq
 k(\|\psi\|^2+\scalprod{\psi}{x^2\psi})\leq k\|\Lambda_p^{1/2}\psi\|^2.
\end{equation*}
Hence $C_3$ satisfies \fer{nc2} and the proof of Proposition \ref{Cprop} is
complete. \hfill $\blacksquare$

\subsection{Proof of Proposition \ref{fgrprop}}
\label{propproofsection}
We consider first the case when \fer{FGRC} holds. From $\Pi I\Pi=0$ we have
$\Pi I\repsilonbar^2 I\Pi=\Pi I\repsilon^2 I\Pi$. We recall that $P_0$ is the
projection onto the kernel of $L_p$ and $\Pbar_0=\bbbone- P_0$ is given by
\begin{equation}
\Pbar_0=\sum_{{m,n\in{\cal M}}\atop{E(m)\neq E(n)}}p_m\otimes p_n + p_d\otimes p_c
+p_c\otimes p_d +p_c\otimes p_c,
\label{a1}
\end{equation}
where $p_d$ and $p_c$ are the projections onto the discrete and continuous
subspaces corresponding to $H_p$. One can see that 
\begin{equation*}
\epsilon \Pi I\repsilon^2 P_0 I\Pi\rightarrow 0, \ \ \epsilon \Pi I\repsilon^2 
\sum_{{m,n\in{\cal M}}\atop{E(m)=E(n)}}(p_m\otimes p_n) I\Pi\rightarrow 0,
\end{equation*}
as $\epsilon\rightarrow 0$, that $\Pi  I\repsilon^2 (p_c\otimes p_c)I\Pi=0$,
and that $\Pi I\repsilon^2 (p_c\otimes p_d) I\Pi=\Pi I\repsilon^2(p_d\otimes
p_c)I\Pi$. From formula \fer{liouvillian} for the interaction, we obtain the bound 
\begin{eqnarray*}
\Pi I\repsilonbar^2 I\Pi&\geq &\Pi I\repsilon^2(p_c\otimes p_d) I\Pi\\
&=&\sum_{{m,n,m'\in{\cal M}}\atop{E(m)=E(n)=E(m')}} \sum_{\alpha,\alpha'}
(p_m\otimes p_n\otimes P_\Omega)\left\{ G_\alpha\otimes\bbbone_p\otimes
  a(\tau_\beta(g_\alpha))\right\}\\
&&\times \frac{p_c\otimes p_d\otimes\bbbone_f}{L^2_0+\epsilon^2} \left\{
  G_{\alpha'}\otimes\bbbone_p \otimes a^*(\tau_\beta(g_{\alpha'}))\right\}
(p_{m'}\otimes p_n\otimes P_\Omega).
\end{eqnarray*}
We write $a(\tau_\beta(g_\alpha))=\int_{\r\times
  S^2}\overline{\tau_\beta(g_\alpha)}(u,\Sigma) a(u,\Sigma)$ and use the pull
  through formula $a(u,\Sigma) L_0=(L_0+u)a(u,\Sigma)$ and obtain
\begin{eqnarray}
\Pi I\repsilonbar^2 I\Pi &\geq& \!\!\!\!\!\!\!
\sum_{{m,n,m'\in{\cal M}}\atop{E(m)=E(n)=E(m')}}
\sum_{\alpha,\alpha'}\int_{-\infty}^{E(m)} du\int_{S^2} d\Sigma
\frac{u^2}{e^{-\beta u}-1}
g_\alpha(-u,\Sigma)\overline{g_{\alpha'}}(-u,\Sigma) \nonumber\\
&&\times\left( p_mG_\alpha\frac{p_c}{(H_p-E(m)+u)^2+\epsilon^2} G_{\alpha'}
  p_{m'}\right)\otimes P_n\otimes P_\Omega,
\label{a2}
\end{eqnarray}
where we recall that $E(m)$ is the eigenvalue of $H_p$ corresponding to the
mode $m$. We have dropped the integration over the values $u\geq E(m)$ because
$\epsilon ((H_p-E(m)+u)^2+\epsilon^2)^{-1}\rightarrow \delta(H_p-E(m)+u)$ as
$\epsilon\rightarrow 0$, hence $u=-H_p+E(m)\leq E(m)$. Recalling the definition of
$F$, see before \ref{T}, and making the change of variable $u\mapsto -u$ in the
integral, we arrive at 
\begin{eqnarray}
\Pi I\repsilonbar^2 I\Pi&\geq &
\sum_{{m,n,m'\in{\cal M}}\atop{E(m)=E(n)=E(m')}}
\int_{-E(m)}^\infty du\int_{S^2} d\Sigma
\frac{u^2}{e^{\beta u}-1}\label{a3}\\
&&
\times
\left( p_m
  F(u,\Sigma)\frac{p_c}{(H_p-E(m)-u)^2+\epsilon^2}
  F(u,\Sigma)^*p_{m'}\right)\otimes p_n\otimes P_\Omega.
\nonumber
\end{eqnarray}
The projection $p(E)$ onto the eigenspace corresponding to an eigenvalue $E$
of $H_p$ is given by $\sum_{m\in{\cal M}:\ E(m)=E} p_m$ and we use  
\begin{equation*}
\sum_{{m,n,m'\in{\cal M}}\atop{E(m)=E(n)=E(m')}} =\sum_{E\in\sigma_p(H_p)}
\sum_{{m\in{\cal M}}\atop{E(m)=E}}
\sum_{{n\in{\cal M}}\atop{E(n)=E}}
\sum_{{m'\in{\cal M}}\atop{E(m')=E}}
\end{equation*}
to arrive at 
\begin{eqnarray}
\lefteqn{
\Pi I\repsilonbar^2 I\Pi\geq  \sum_{E\in\sigma_p(H_p)}\int_{-E}^\infty
du\int_{S^2}d\Sigma \frac{u^2}{e^{\beta u}-1}}\label{a4} \\
&&\left( p(E)
  F(u,\Sigma)\frac{p_c}{ (H_p-E-u)^2+\epsilon^2} F(u,\Sigma)^*p(E)\right)
\otimes p(E)\otimes P_\Omega.
\nonumber
\end{eqnarray}
The desired bound \fer{fgrc1} now follows from \fer{FGRC} and \fer{gamma}. \\
\indent
The case when \fer{FGRC'} holds and $\gamma$ is given by \fer{gamma'} is done
similarly. \hfill $\blacksquare$

\end{document}